\documentclass[%
superscriptaddress, 
preprint,
 amsmath,amssymb,
 aps,
]{revtex4-1}

\usepackage{graphicx}
\usepackage{dcolumn}
\usepackage{bm}
\usepackage{braket}
\usepackage{float}
\usepackage{amsmath}

\begin{document}

\newcommand{\smb}{SmB$_{6}$}
\newcommand{\kt}{k_B T}

\title{Bulk transport paths through defects in floating zone and Al flux grown SmB$_6$}

\author{Yun Suk Eo}
    \email[]{eohyung@umich.edu}
    \thanks{These authors contributed equally.}
    \affiliation{Department of Physics, University of Michigan, Ann Arbor, MI 48109 USA}
    \affiliation{Maryland Quantum Materials Center and Department of Physics, University of Maryland, College Park, Maryland, USA}
\author{Alexa Rakoski}
    \email[]{ralexa@umich.edu}
    \thanks{These authors contributed equally.}
    \affiliation{Department of Physics, University of Michigan, Ann Arbor, MI 48109 USA}
\author{Shriya Sinha}
    \affiliation{Department of Physics, University of Michigan, Ann Arbor, MI 48109 USA}
\author{Dmitri Mihaliov}
    \affiliation{Department of Physics, University of Michigan, Ann Arbor, MI 48109 USA}
\author{Wesley T. Fuhrman}
    \affiliation{Maryland Quantum Materials Center and Department of Physics, University of Maryland, College Park, Maryland, USA}
    \affiliation{Institute for Quantum Matter and Department of Physics and Astronomy, The Johns Hopkins University, Baltimore, MD 21218 USA}
\author{Shanta R. Saha}
    \affiliation{Maryland Quantum Materials Center and Department of Physics, University of Maryland, College Park, Maryland, USA}
\author{Priscila F. S. Rosa}
    \affiliation{Los Alamos National Laboratory, Los Alamos, NM 87545 USA}
\author{Zachary Fisk}
    \affiliation{Department of Physics and Astronomy, University of California Irvine, Irvine, CA 92697 USA}
\author{Monica Ciomaga Hatnean}
    \affiliation{Department of Physics, University of Warwick, Coventry, CV4 7AL, UK}
\author{Geetha Balakrishnan}
    \affiliation{Department of Physics, University of Warwick, Coventry, CV4 7AL, UK}
\author{Juan R. Chamorro}
    \affiliation{Institute for Quantum Matter and Department of Physics and Astronomy, The Johns Hopkins University, Baltimore, MD 21218 USA}
    \affiliation{Department of Chemistry, The Johns Hopkins University, Baltimore, MD 21218 USA}
\author{Seyed M. Koohpayeh}
    \affiliation{Institute for Quantum Matter and Department of Physics and Astronomy, The Johns Hopkins University, Baltimore, MD 21218 USA}
    \affiliation{Department of Materials Science and Engineering, The Johns Hopkins University, Baltimore, MD 21218 USA}
\author{Tyrel M. McQueen}
    \affiliation{Institute for Quantum Matter and Department of Physics and Astronomy, The Johns Hopkins University, Baltimore, MD 21218 USA}
    \affiliation{Department of Chemistry, The Johns Hopkins University, Baltimore, MD 21218 USA}
     \affiliation{Department of Materials Science and Engineering, The Johns Hopkins University, Baltimore, MD 21218 USA}
 
\author{Boyoun Kang}
    \affiliation{Department of Materials Science and Engineering, Gwangju Institute of Science and Technology, Gwangju 61005 Korea}
\author{Myung-suk Song}
    \affiliation{Department of Materials Science and Engineering, Gwangju Institute of Science and Technology, Gwangju 61005 Korea}
\author{Beongki Cho}
 \affiliation{Department of Materials Science and Engineering, Gwangju Institute of Science and Technology, Gwangju 61005 Korea}
 
\author{Michael S. Fuhrer}
    \affiliation{Monash University, Melbourne, Victoria 3800, Australia}
    \affiliation{ARC Centre of Excellence in Future Low-Energy Electronics Technologies, Monash University, Victoria 3800 Australia}
 
\author{Johnpierre Paglione}
    \affiliation{Maryland Quantum Materials Center and Department of Physics, University of Maryland, College Park, Maryland, USA}
    \affiliation{Canadian Institute for Advanced Research, Toronto, Ontario M5G 1Z8, Canada}
    
\author{\c{C}a\u{g}l{\i}yan Kurdak}
    \affiliation{Department of Physics, University of Michigan, Ann Arbor, MI 48109 USA}

\date{\today}

\begin{abstract}
We investigate the roles of disorder on low-temperature transport in SmB$_6$ crystals grown by both the Al flux and floating zone methods. We used the inverted resistance method with Corbino geometry to investigate whether low-temperature variations in the standard resistance plateau arises from a surface or a bulk channel in floating zone samples. The results show significant sample-dependent residual bulk conduction, in contrast to smaller amounts of residual bulk conduction previously observed in Al flux grown samples with Sm vacancies. We consider hopping in an activated impurity band as a possible source for the observed bulk conduction, but it is unlikely that the large residual bulk conduction seen in floating zone samples is solely due to Sm vacancies. We therefore propose that one-dimensional defects, or dislocations, contribute as well. Using chemical etching, we find evidence for dislocations in both flux and floating zone samples, with higher dislocation density in floating zone samples than in Al flux grown samples. In addition to the possibility of transport through one-dimensional dislocations, we also discuss our results in the context of recent theoretical models of SmB$_6$.
\end{abstract}

\maketitle

\section{Introduction}
SmB$_6$ is the oldest known Kondo insulator (KI), \cite{aeppli, riseborough00} in which strong correlations between the $f$ and $d$ electrons lead to the opening of a small hybridization gap at the Fermi energy below about 100 K. The initial narrow-gap picture resulting from hybridization of $f$- and $d$- bands was proposed by N. Mott in 1974. \cite{mott74} Since then, many reports have elaborated on this picture, but lingering problems persisted. Experimentally, one problem was the resistivity saturation below about 4 K which could not be explained with impurities or a minimum conductivity model. \cite{menth, allen} Theoretically, the Kondo hybridization that opened the gap had parity violation at the high symmetry points. \cite{martin1979theory}

After nearly 40 years, shortly after the advent of three-dimensional topological insulators (TI) in bismuth-based materials, these two lingering mysteries of SmB$_6$ were revisited and arguably solved by introducing a topological band inversion. When the hybridization gap forms from two bands of opposite parities, it leaves a surface conduction channel that is topologically protected. \cite{dzero10,takimoto, dzero12} Transport experiments confirmed that the surface is conductive while the bulk is insulating. \cite{wolgast13, kim13, kim14} Experimental evidence supporting the TI proposal was also obtained via methods including angle resolved photoemission spectroscopy (ARPES), \cite{neupane, jiang, nxu14, denlingerlinked} point contact spectroscopy, \cite{zhang13} scanning tunneling microscopy, \cite{rossler16, pirie} and inelastic neutron scattering. \cite{fuhrman15} Conversely, some reports favor a non-topological explanation for surface conduction in SmB$_6$. \cite{hlawenka, hermann, frantzeskakis} 

Samples used in modern studies are grown either by the aluminum flux method or the optical floating zone method. Single crystals can be grown below their melting point by the Al flux method, which may enhance the stoichiometry of the target sample by preventing vaporization of Sm at high temperatures. \cite{phelan16} Flux grown samples are small (a few mm in each direction) and can contain inclusions of the flux. \cite{canfield} In contrast, floating zone samples are grown at or above the melting point, and the high temperatures used can introduce defects due to thermal stresses \cite{koohpayeh} or through vaporization of Sm. \cite{phelan16} Floating zone samples are quite large (a few cm long) and are uncontaminated by flux. In general, characterization methods like powder x-ray diffraction show no obvious difference between samples grown by the two methods. \cite{hatnean, phelan16} However, there appears to be a clear difference in the experimental results, especially at low temperatures, when comparing Al flux and floating zone grown samples. ARPES results that find evidence for a trivial surface in SmB$_6$ were performed on floating zone samples, \cite{hlawenka} while some of the most compelling evidence for a topological surface comes from flux grown samples. \cite{nxu14}

De Haas van Alphen (dHvA) quantum oscillations were also used to search for TI states, \cite{liscience, tan, xiang17, hartstein} but instead they revealed deeper mysteries about the bulk of SmB$_6$ and the origin of the low temperature behavior. Reports finding evidence for a 2D surface \cite{liscience, xiang17} used flux grown samples, but these results have also been attributed to aluminum inclusions in the samples. \cite{thomas19} Tan, $et~al.$ \cite{tan} observe bulk quantum oscillations in floating zone grown SmB$_6$ at low temperatures when the bulk gap is opened. In addition, heat capacity and thermal conductivity on some floating zone grown samples show a large residual density of states in the $T\rightarrow 0$ limit, which could imply that charge-neutral fermions exist in the bulk. \cite{hartstein} However, flux-grown samples have never shown evidence for charge-neutral quasiparticles in the bulk. \cite{xu16, boulanger} These and other subsequent experimental and theoretical studies attempting to resolve the bulk dHvA result take two opposite approaches. Bulk quantum oscillations could be intrinsic to SmB$_6$, for example due to charge-neutral quasiparticles, or they could have an extrinsic origin, for example, pockets of an unknown metallic phase.

Much of the theoretical work on quantum oscillations has focused on a possible intrinsic origin. Some of these scenarios have included oscillations by excitonic states \cite{knolle17, chowdhury} or a Majorana fermion band that breaks gauge symmetry. \cite{baskaran, erten17} Others have proposed breakdown of the gap under magnetic field, \cite{erten16, riseborough17} or ways for oscillations to occur in gapped systems based on the unhybridized band structure or as an effect of the band edges. \cite{knolle15, zhang16} 

The other possibility is that the quantum oscillations have an extrinsic origin from disorder or impurities. In the presence of generic short-range disorder, states from the conduction and valence band could spill into the gap, which could be responsible for the oscillations. \cite{shen, harrison} Alternatively, natural magnetic impurities could be responsible for the excess heat capacity at low temperatures. \cite{fuhrman18} These local moments in the lattice would be screened, and the amount of screening, and thus the magnetization, would oscillate in magnetic field. \cite{fuhrman182} Still another report focused on nonmagnetic impurities, which were found to form a deep impurity band as in a metal as well as an in-gap band, \cite{abele} and another proposal revisited the idea of in-gap impurity states. \cite{skinner} Historically, hydrogenic in-gap impurity states as are found in doped semiconductors were proposed in SmB$_6$. This model is unjustified in SmB$_6$, \cite{rakoski17}, one reason being that the standard hydrogenic impurity model relies on a parabolic band structure. Instead, Ref. \cite{skinner} shows that the hybrid band structure of SmB$_6$ has its own model of hydrogen-like in-gap impurity states. Interestingly, the density of defects required for an insulator-to-metal transition is orders of magnitude higher than the required density for the same in parabolic semiconductors. 

These proposals for extrinsic sources of quantum oscillations have experimental consequences well beyond bulk quantum oscillations. For example, in transport, the nodal semimetal scenario would imply linear-in-T behavior in the bulk resistivity at low temperatures. \cite{harrison} In the presence of hydrogenic-like impurities, low-temperature bulk resistivity would be dominated by an activated term corresponding to hopping in the impurity band; the activation energy would be different from the one arising from the Kondo gap. \cite{skinner} In our previous work, we used the inverted resistance method to find the bulk resistivity even when the surface channel dominates below about 4 K. We found that SmB$_6$ grown by the aluminum flux method shows a continuous exponential rise in resistivity of nearly 10 orders of magnitude from 40 K to 2 K. \cite{eo18} Samples grown with Sm deficient off-stoichiometry still showed an exponential rise of 7-8 orders of magnitude, but at about 2 K they reveal a bulk saturation distinct from the surface channel. \cite{eo19} 

We previously argued that the resistivity values of this newly discovered bulk channel at low temperatures are extremely high. In fact, such resistivity saturation after a high magnitude increase is only seen in ultra clean semiconductors. \cite{debye, chapman} This would correspond to a tiny conduction channel, which does not help resolve the question of quantum oscillations. Nevertheless, understanding the origin of this low-temperature bulk conduction is important for understanding the unique role of disorder in SmB$_6$. Previously, we speculated that the mysterious third channel could be conduction through one-dimensional defects, or dislocations, that are topologically protected. \cite{eo19} 

Dislocations have been studied extensively in semiconductor thin films such as GaN, where they are a significant source of scattering in electronic devices and provide recombination sites in optoelectronic devices. \cite{ourmazd} In thin films, dislocations form during growth, especially at the interface between a substrate and a film with different lattice constants. This lattice mismatch between the two materials strains the layer, leading to the formation of dislocations. \cite{hullbean} The density of dislocations present in the film is related to the difference in lattice constants between the substrate and the film, with lower dislocation density corresponding to more closely matched lattice constants. Dislocations are also present in crystals. They can form from internal stresses in the growth, especially stresses due to thermal fluctuations, local impurities in the growth, or even vibrations in the environment. \cite{hullbean} Impurities in the growth can provide nucleation sites where dislocations start to form, and high temperatures used in the growth can compound the effect of internal stresses as well. \cite{hull} Additionally, dislocations can extend from a seed crystal containing dislocations to new growth based on that seed. \cite{hull} However, not much is known about dislocations in topological materials. Previously, dislocations in Bi-based topological thin films have been shown to create unwanted bulk current paths. \cite{hamasaki, ran} Dislocations in SmB$_6$ would be especially interesting to study in light of the previous report of a truly insulating bulk in the dc limit. 

The level of disorder can generally be measured in transport via the mobility. At low temperatures, measurements of mobility in SmB$_6$ are not straightforward. Experimental reports - ARPES, dHvA oscillations, and transport - disagree on the order of magnitude of the mobility, which ranges from about 10 cm$^2$/V$\cdot$s in transport \cite{luo15} to about 1000 cm$^2$/V$\cdot$s in quantum oscillations; \cite{liscience} surface preparation can even affect the extracted mobility. \cite{eo20} A recent study also shows that the two proposed topological surface channels would have very different mobilities, \cite{eo20} and accounting for a disorder-based channel could be an additional challenge. 

Detailed transport results have also shown that the resistivity saturation at low temperature is non-universal. \cite{phelan16} Al-flux grown samples generally yield resistivity with temperature-independent plateaus. \cite{cooleyprl, sluchanko99, luo15} Floating zone samples are generally less consistent and can behave similarly to flux-grown samples, show temperature-dependent behavior, or even a step-like behavior. \cite{flachbart012, gabani15, phelan16} An open question is whether these differences in behavior are due to different surface characteristics or bulk characteristics. If the differences are due to surface characteristics, the low-temperature bulk behavior should be similar between both types of samples. But, if bulk characteristics differ at low temperatures, this would be reflected in inverted resistance measurements. 

In this work, we perform inverted resistance measurements on a Corbino disk geometry on floating zone grown samples. We find that these samples all demonstrate bulk conduction at lowest temperatures, but with significant sample-to-sample variation. In combination with previous results identifying a low-temperature bulk conduction channel on flux-grown samples of different defect levels, \cite{eo19} we discuss the possible origins of this new bulk channel in the context of recent impurity models and dislocations. To expand experimental understanding of the role of disorder, we also perform chemical etching to verify the presence of dislocations in our flux and floating zone grown samples. The wide variation in our low-temperature results depending on the sample used suggests that many discrepancies in experimental reports on SmB$_6$ may have an extrinsic origin.

\section{Inverted resistance on a Corbino disk}
To investigate the origin of different low-temperature behavior in SmB$_6$, we used a recently developed method called inverted resistance. This method can distinguish whether the resistance originates from the bulk or the surface and allows us to find the bulk resistivity even if the surface conduction overwhelms the bulk. \cite{eo18} We briefly illustrate the method. Consider the simple case where the resistance only depends on the bulk resistivity. This is when the bulk conduction overwhelms the surface conduction or the surface conduction does not exist at all. In this case, the electric current flows only through an isotropic bulk, and the resistance is proportional to the bulk resistivity, $\rho_{b}$:
\begin{equation}
	R = C_{b} \rho_{b},
	\label{Eq:BulkResistance}
\end{equation}
where $C_{b}$ is a prefactor that is determined by the geometry of the sample and the position of the electrodes. The resistance measurements from a different selection of electrodes will only change the value $C_{b}$. Those different resistance measurements will have the same temperature dependence, originating from $\rho_{b}$, and therefore the $R$ vs. temperature and $\rho_b$ vs. temperature curves will have the same shape and be parallel.

If the two resistances are not parallel to each other as a function of temperature, Eq. (1) cannot be used. One possible reason for not being parallel is when disorder in the crystal creates a large inhomogeneity and therefore the temperature dependence of the bulk resistivity is not global. A more dramatic case is when an extra conduction channel is present, for example a surface conduction channel. This is indeed the case for SmB$_6$ below 4 K, and the resistance measurement can be explained simply by the following. The bulk resistivity increases exponentially with inverse temperature while the surface sheet resistance changes only very moderately. At low enough temperatures, the bulk conduction is so low that it becomes overwhelmed by the surface conduction. In this temperature regime, the measured resistance is proportional to the sheet resistance, $R_{s}$:
\begin{equation}
	R = C_{s} R_{s},
	\label{Eq:SurfaceResistance}
\end{equation}
where $C_{s}$ is a prefactor that is determined by the geometry of the surface and the position of the electrodes. 

In the case of both surface and bulk conduction, a different type of resistance measurement can be used. Here, the current flows inside a Corbino disk geometry and the voltage is measured exterior to that disk. This method is known as inverted resistance, and the details can be found in Ref.~\cite{eo18}. The inverted resistance, $R_{\textrm{Inv}}$, now depends on both $R_{s}$ and $\rho_{b}$:
\begin{equation}
	R_{\textrm{Inv}} = C_{\textrm{Inv}} \frac{R_{s}^2}{\rho_{b}},
	\label{Eq:InvResistance}
\end{equation}
We use Eqs.~\ref{Eq:BulkResistance},~\ref{Eq:SurfaceResistance}, and~\ref{Eq:InvResistance} to analyze our resistance measurement. 
$R_{\textrm{Inv}}$ is particularly useful because it allows us to access $\rho_{b}$ at temperatures where surface conduction dominates. In the case where the bulk resistivity exhibits ideal activated behavior, $\rho(T) \propto \exp{(E_{a}/T)}$, $R_{Inv}$ will also follow the inverse temperature dependence ($\propto \exp{(-E_{a}/T)}$). This ideal relation surprisingly holds true for the case of stoichiometric flux-grown SmB$_{6}$ case \cite{eo19} However, this temperature dependence in the resistance can be interrupted if a second bulk conduction channel, exists.

\begin{figure}
\includegraphics[scale=0.7]{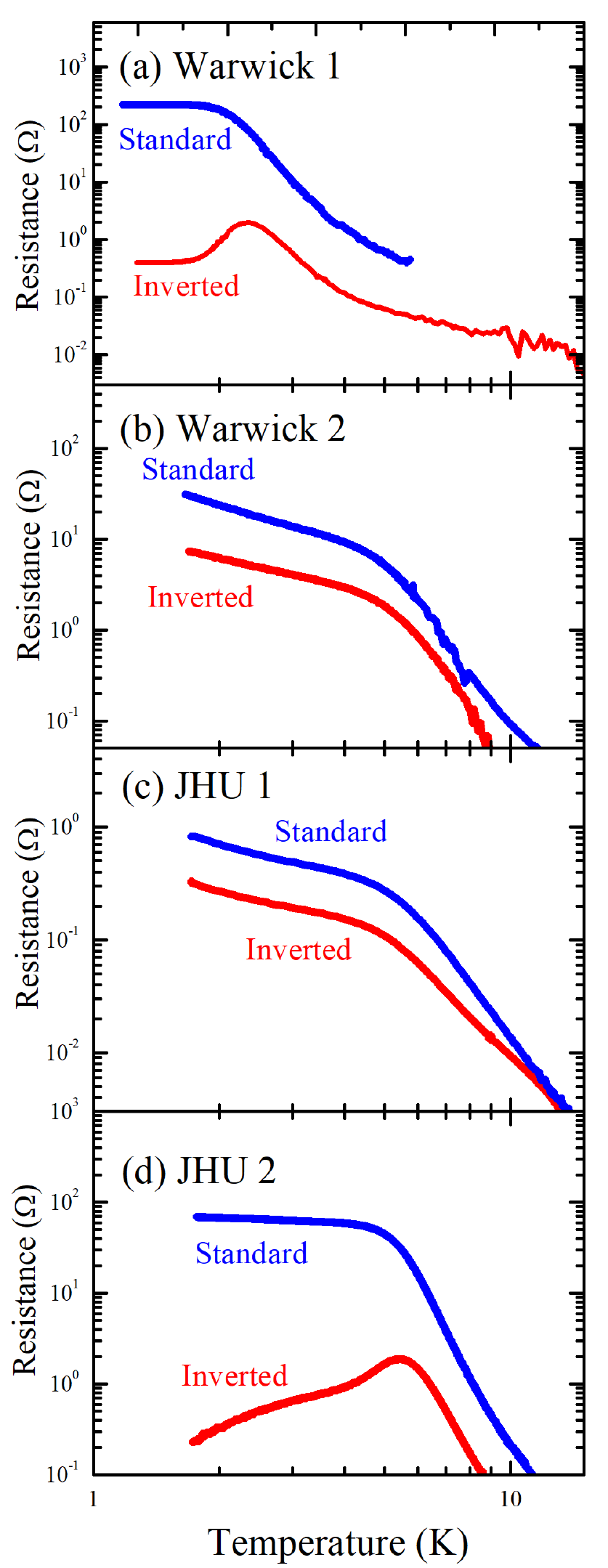}
\caption{\label{Fig:Yun_InvertedRfig} Inverted resistance measurement of four floating-zone grown SmB$_6$ samples.}
\end{figure}

\section{Results}
Four floating zone samples were prepared for the inverted resistance measurement. Details of the samples' origins and growth methods can be found in Table~\ref{tab:sampledetails}. All samples were polished with grits down to 0.3 $\mu$m. The Corbino disks were patterned by photolithography. Ti/Au was deposited on the samples using e-beam evaporation and later a lift-off process with acetone was used to define the pattern and electrodes. We used gold wires to make electrical connection from the electronics to the sample, and attached them using either silver paste or wire bonding. 

\begin{table}
\centering
\begin{ruledtabular}
\begin{tabular}{l c c c c}
Sample & Growth details & Starting powder origin & Reference \\
\hline
Warwick 1 & Standard growth & Alfa Aesar & \cite{hatnean} \\
Warwick 2 & Standard growth & American Elements & \cite{hatnean} \\
JHU 1  & Standard growth & Testbourne Ltd. & \cite{phelan16} \\
JHU 2 & Doubly-isotope enriched, Sm deficient & Alfa Aesar & \cite{phelan16} \\
\end{tabular}
\end{ruledtabular}
\caption{\label{tab:sampledetails}Details of the four floating zone samples measured.}
\end{table}

Fig.~\ref{Fig:Yun_InvertedRfig} shows the resistance vs. temperature of all four samples. The blue curves are the standard resistance measurements and the red curves are the inverted resistance measurements. In the standard measurements (blue curves), all four samples show a change in slope around 4 K that would conventionally be regarded as a surface plateau. Fig.~\ref{Fig:Yun_InvertedRfig}~(a) and (d) show little to no temperature dependence in the standard measurement below about 4 K, especially compared to Fig.~\ref{Fig:Yun_InvertedRfig}~(b) and (c). Using the inverted curves (red), we can determine whether these ``plateaus" arise from surface or bulk conduction channels. We see dramatic differences in the inverted resistance results for each sample. Fig.~\ref{Fig:Yun_InvertedRfig}~(a) shows a resistance that drops and saturates, similar to the non-stoichiometric flux growths observed previously. \cite{eo19} The inverted resistance measurement shown in Fig.~\ref{Fig:Yun_InvertedRfig}~(d) has a feature similar to Fig.~\ref{Fig:Yun_InvertedRfig}~(a) but also a moderate drop at lower temperatures. In Fig.~\ref{Fig:Yun_InvertedRfig}~(b) and (c), we see a temperature dependence that is close to parallel to the standard measurement. 

\begin{figure}
\includegraphics[scale=1]{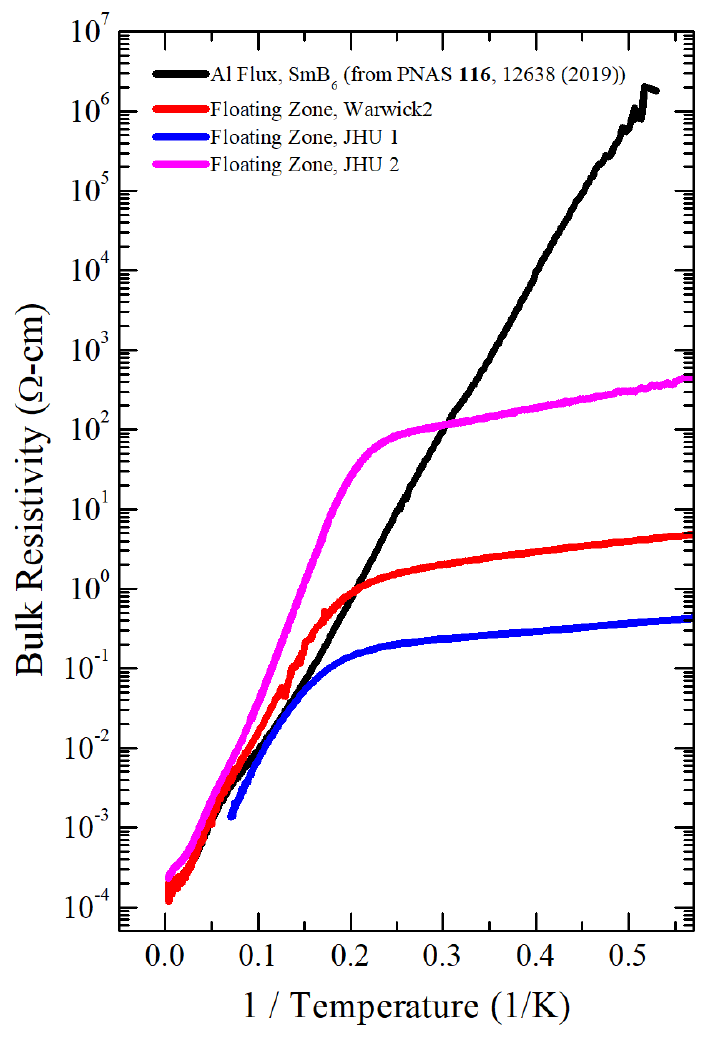}
\caption{\label{Fig:Yun_bulkresistivity} Bulk resistivity conversion from Fig.~\ref{Fig:Yun_InvertedRfig} measurements.}
\end{figure}

We convert the measured resistances to bulk resistivity in Fig.~\ref{Fig:Yun_bulkresistivity}, except for the sample Warwick 1 (Fig.~\ref{Fig:Yun_InvertedRfig}(a)) which was complicated to convert due to the Corbino disks being placed on different crystal planes. We compare the results from this study to the previously reported result on a stoichiometric flux-grown sample (black trace). \cite{eo19} We find that the remaining three floating zone-grown samples have a significant slope change in bulk resistivity which indicates that the intrinsic exponential temperature dependence in bulk resistivity is interrupted. That is, another bulk conduction mechanism is present in these samples in addition to the standard mechanism responsible for activated behavior.  

\section{Discussion}
Figs.~\ref{Fig:Yun_InvertedRfig} and \ref{Fig:Yun_bulkresistivity} show that the four floating zone samples presented here have non-negligible bulk conduction, but the characteristics of the bulk conduction differ by sample. In Fig.~\ref{Fig:Yun_InvertedRfig}~(b) and~(c) the standard and inverted resistance curves are parallel to one another, demonstrating that these samples are bulk conductors and can be described by Eq.~\ref{Eq:BulkResistance}. In terms of transport, this means that either the surface conduction is nonexistent or that the bulk conduction channel dominates. 

Many researchers have proposed impurities as a possible origin for residual bulk conduction in SmB$_6$. Most point defects come from the starting material, as rare earth elements are notoriously difficult to purify. In our samples, the starting materials were sourced from different companies, so the purity of the Sm used in the growth may differ among samples. One way to reduce rare earth impurities is via isotopic purification to Sm-154. The only remaining rare-earth impurity is Gd-154, which is magnetic. \cite{fuhrman18} Gd impurities have been studied previously, beginning with the observation that the substitution of even 1{\%} Gd could dramatically change the electrical properties of SmB$_6$. \cite{geballe} Later, a Gd doped sample was used to test the TI hypothesis by searching for time reversal symmetry breaking below 4 K. \cite{kim14} Gd impurities have also been shown to increase the residual heat capacity at low temperatures \cite{fuhrman18} and have been suggested as an avenue for screening of the Kondo effect at low temperatures. \cite{fuhrman19} Debate is ongoing about the role of Gd impurities; recent reports have suggested that it is not responsible for bulk dHvA oscillations. \cite{hartstein20} However, the local environment of Gd impurities is metallic even at very low concentrations, and at higher concentrations this could lead to percolation through the sample in transport measurements. \cite{souza20}

Results from the isotopically purified sample shown in Fig.~\ref{Fig:Yun_InvertedRfig}~(d) have different features in the inverted resistance curve compared to the non-purified samples (Figs.~\ref{Fig:Yun_InvertedRfig}~(b) and (c)). Since the standard and inverted curves are not parallel, this sample does not have dominant bulk conduction, but it may have parallel surface and bulk channels. Since the sample is isotopically pure, the bulk channel could come from the remaining Gd impurities, Sm vacancies introduced during growth, or both.

A general model of impurities used in semiconductors and other materials is the effective mass approximation, where the impurity is treated hydrogenically, with an effective Bohr radius and binding energy. \cite{kohn} In SmB$_6$, the conditions for standard hydrogenic impurites are not satisfied when the model for semiconductors is used. \cite{rakoski17} However, the model introduced by B. Skinner in Ref. \cite{skinner} demonstrated that the effective mass approximation can be modified for SmB$_6$. In the Skinner model, the quadratic potential used in the original treatment of the effective mass approximation \cite{kohn} is swapped for the Mexican hat type potential seen in SmB$_6$. New conditions for the effective radius and binding energy of the impurity state are determined. The total dc conductivity in the presence of these new impurity states is also derived and found to be a combination of the standard activated behavior with activation energy $E_1$ and an activated hopping term with activation energy $E_3$, \cite{skinner}
\begin{equation}
\sigma(T) = \sigma_1 \exp{\bigg( \frac{E_1}{k_B T} \bigg)} + \sigma_3 \exp{\bigg( \frac{E_3}{k_B T} \bigg)}. 
\end{equation}
This type of impurity band could be present in all samples and could describe both magnetic and nonmagnetic point impurities, including Sm vacancies. Since the addition of Sm vacancies to flux grown samples has been shown to induce bulk conductivity, \cite{eo19} some portion of the residual bulk conductivity seen in this work in floating zone samples could also be due to Sm vacancies as described above. However, the magnitude of the bulk conduction seen in the inverted measurements is much greater in all the floating zone samples, including the isotopically purified sample, so it is unlikely that vacancies or impurities alone could be the origin. The possibility of hopping conduction and even insulator-to-metal transition by heavily doped foreign magnetic impurities will be discussed elsewhere in our future work.

Another possibility for the source of the residual bulk conduction is one-dimensional defects, or dislocations. As discussed earlier, a mismatch in lattice constant between a substrate and a semiconductor thin film can lead to the formation of dislocations which terminate on the surface of the film. \cite{hullbean} The density of dislocations in a film can be estimated by the change in lattice parameter, $n_{\textrm{dis}} = |1/a_1^2 - 1/a_2^2|$, where $a_1$ and $a_2$ are the lattice parameters on the two surfaces of the film.

In SmB$_6$, one study reports a change in lattice constant over the length of a floating zone sample. \cite{phelan16} Unlike the case of thin films, to the best of our knowledge there is no literature describing how to estimate dislocation density in bulk crystals where there is a variation in lattice constant. Here, we introduce a new method to understand the formation of dislocations in bulk materials. Later, we use our model to estimate the dislocation density in SmB$_6$ samples. We consider a floating zone sample as its size allows for more variation of lattice parameter in the crystal compared to a flux grown sample, but dislocations are still expected to be present in flux grown crystals. 

\begin{figure}
\includegraphics[scale=.65]{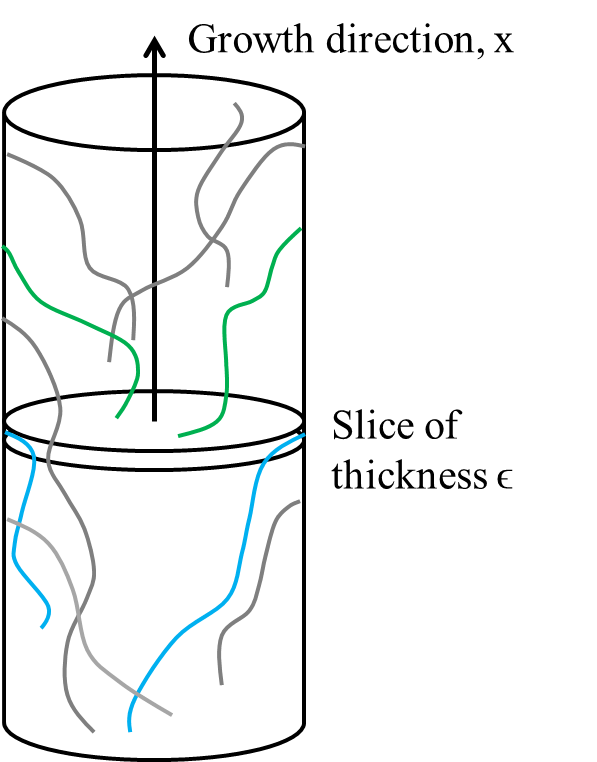} 
\caption{\label{fig:dislocationsketch} Sketch of dislocations in a floating zone crystal. Some dislocations (green) initiate within a slice of thickness $\epsilon$, and others (blue) terminate within that slice. The remaining dislocations (gray) form and terminate elsewhere in the sample.}
\end{figure}

Generally, dislocations that form in films terminate on the surface of the film. In thin films, the dislocations form in the growth direction and terminate on the top surface of the film. In crystals, however, dislocations do not have to form and terminate only along the direction of growth; dislocations could also terminate on the side surfaces of the crystal, as shown in Fig.~\ref{fig:dislocationsketch}. Thus, the dislocation density in crystals is expected to depend both on the change in lattice constant in the direction of growth and on the size (radius) of the crystal. 

To estimate the dislocation density, we model the floating zone rod as forming from a series of thin slices of thickness $\epsilon$ as the molten zone passes through the furnace. In analogy with the equation above for semiconductor thin films, the total number of dislocations that nucleate within the slice is
\begin{equation}
\label{eq:dislocations-lhs}
\left| \frac{1}{(a(x))^2} - \frac{1}{(a(x+\epsilon))^2} \right| (\pi r^2)
\end{equation}
where $a(x)$ is the lattice parameter at a location $x$ along the growth direction and $r$ is the radius of the crystal. A sketch of these is shown in green in Fig.~\ref{fig:dislocationsketch}. All the dislocations that form in this slice must terminate somewhere on the surface of the sample, whether on the sides or the ends. To account for the dislocation terminating on the sides, we introduce an angle $\theta_{\textrm{d}}$ which the dislocation makes with respect to the growth direction. Then, the number of dislocations that terminate within a slice (shown in blue in Fig.~\ref{fig:dislocationsketch}) is related to the surface area of the slice, the dislocation density ($n_{\textrm{dis}}$), and $\theta_{\textrm{d}}$ by
\begin{equation}
\label{eq:dislocations-rhs}
(2 \pi r \epsilon) n_{\textrm{dis}} \cos \theta_{\textrm{d}}.
\end{equation}
Here, $\cos \theta_{\textrm{d}} = 1$ would correspond to all dislocations oriented along the growth direction. We expect $cos \theta_{\textrm{d}} < 1$ in an actual sample, since dislocations are expected to terminate randomly on the surface but form with the growth of the rod. Since all the dislocations that formed must terminate, Eqs.~\ref{eq:dislocations-lhs} and \ref{eq:dislocations-rhs} are equal. Expanding Eq. \ref{eq:dislocations-lhs} as a Taylor series, we calculate that the estimated dislocation density is
\begin{equation}
\label{eq:dislocation-density}
n_{\textrm{dis}} = \frac{r}{\cos \theta_{\textrm{d}}} \frac{|\nabla a(x)|}{(a(x))^3}
\end{equation}
for the dislocation density. 

We can estimate the dislocation density in the sample with reported change of lattice constant from Ref. \cite{phelan16}. The lattice parameter in that sample was $a_1 = 4.134309$ \AA{} on one end of the crystal and $a_2 = 4.133343$ \AA{} on the other end, 8 cm away. The radius of the crystal was 3 mm. Using these values with Eq. \ref{eq:dislocation-density}, we estimate that the dislocation density in this floating zone sample is $\sim 10^{10}$ cm$^{-2}$.

Dislocations are commonly imaged by preparing samples as for transmission electron microscopy (TEM). However, the estimated density of dislocations we calculated is too small to use this method. Instead, we used chemical etching to reveal points where dislocations terminate on the surface. During etching, material is removed from the area near a crystal defect at a different rate than from the lattice. The etching method allows defects, including dislocations, to be imaged optically. \cite{ourmazd} The ``etch pit" that forms also mirrors the crystal structure of the sample; for SmB$_6$ we expect to see square etch pits. With longer etching time, the size of the etch pits increases and more etch pits start to form, so that the etch pit density observed provides a lower bound on the actual dislocation density. 

\begin{figure}
\includegraphics[scale=.8]{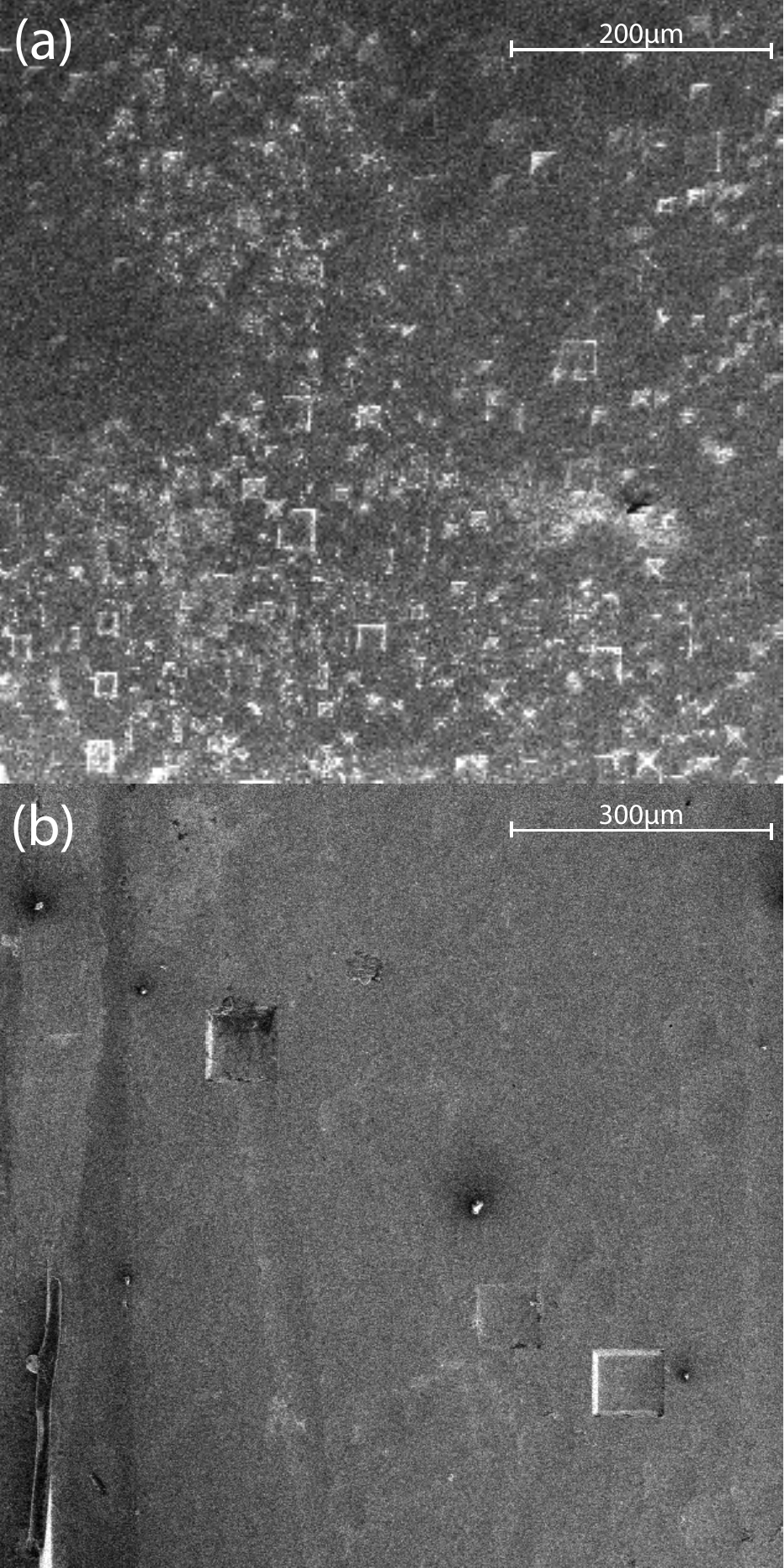} 
\caption{\label{fig:etchpits} Examples of etch pits in (a) a floating zone sample etched for 340 seconds and (b) a flux-grown sample etched for 600 seconds.} 
\end{figure}

We used equal parts nitric and sulfuric acid diluted to 10\% to etch flux grown and floating zone SmB$_6$ crystals. After etching, we observed etch pits using a scanning electron microscope in both types of samples. Examples of etch pits are shown in Fig.~\ref{fig:etchpits}. The floating zone sample shown was etched for 340 seconds and had an etch pit density of $10^{5}$ cm$^{-2}$. The flux grown sample shown was etched for 600 seconds and had an etch pit density of $2\times 10^{3}$ cm$^{-2}$. Even though the flux grown sample was etched longer than the floating zone sample, it has a lower density of etch pits observed, suggesting that the floating zone sample hosts more dislocations than the flux grown sample. In both samples, the locations of the etch pits is nonuniform, which suggests that local inhomogeneities in temperature or stoichiometry, for example, during sample growth are important to the formation of dislocations. In both cases, the observed densities are much lower than the calculated estimate of 10$^{10}$ cm$^{-2}$, but since we imaged immediately after identifying that etch pits were present, our values are lower bounds on the number of dislocations actually present in the samples.

In addition to impurity hopping and dislocations, we briefly consider other theories proposed to explain some of the novel results of SmB$_6$. First, the proposal that SmB$_6$ is a nodal semimetal \cite{shen,harrison} is inconsistent with the low-temperature bulk conduction that we observe, and it does not explain the difference between flux- and floating zone-grown samples. Our inverted resistance curves show two regions of activated behavior (above and below about 4 K) rather than activated behavior above 4 K and linear-in-T behavior below 4 K. Next, in heat transport, excess thermal conduction at low temperature was not found in flux-grown samples, and reports have disagreed about whether thermal conduction is present universally in floating zone samples. \cite{hartstein, xu16, boulanger}. Theories attempting to reconcile these conflicting results have focused on the possibility that the floating zone samples contain charge-neutral excitations and primarily explored their relevance to dHvA oscillations rather than transport. Our data do not provide evidence for charge-neutral excitations, but the low-temperature bulk channel we observe could conduct heat and contribute to dHvA. Even in samples with very few rare earth impurities, \cite{hartstein20} dislocations could still contribute to these effects. A better understanding of the role of dislocations, or more generally, the conduction channel we observe here, will be an intriguing area of further study. 

\section{Conclusion}
In this work, we performed transport measurements on floating zone grown SmB$_6$ using the inverted resistance method. Standard four point and Hall bar geometry resistance results show nonuniversal temmperature dependence below about 4 K, but the origin of this behavior is difficult to pinpoint as both bulk and surface channels are present. The inverted resistance method we used allowed us to characterize the bulk behavior at temperatures at which surface conduction dominates. We found that the four floating zone samples show bulk conduction with characteristics differing by sample. On the other hand, a stoichiometric flux-grown sample (along with other results from Ref.~\cite{eo19}) has a truly insulating bulk, and the introduction of Sm vacancies in flux grown samples was previously shown to induce bulk conduction. 

We discussed various possibilities for the origin of the new conducting channel observed here, as well as the differences between the floating zone results presented here and the flux-grown samples presented in the previous work. \cite{eo19} We especially considered impurities, which could be magnetic, like Gd, or non-magnetic, including defects like Sm vacancies. Our experimental results are consistent with the Skinner model \cite{skinner} for impurity hopping conduction at low temperatures with an activated transport behavior. In addition, we considered one-dimensional defects, or dislocations, extending throughout the sample. We observed a small dislocation density in both flux and floating zone samples via chemical etching, with a larger dislocation density observed in floating zone samples compared to flux grown samples. While this is consistent with the relative amounts of bulk conduction observed in samples grown by each technique, further work to explore the characteristics of the dislocations is needed to verify that they contribute to bulk conduction with a magnitude agreeing with our inverted resistance data. Future work could include characterizing the mobility of the channel or thermal studies of the role of dislocations in SmB$_6$.

\begin{acknowledgments}
A.R. would like to acknowledge funding support from NSF Grant \#DGE-1256260. W.T.F. is grateful to the Schmidt Science Fellows program, in partnership with the Rhodes Trust, for support of this work. M.S.F. and Y.S.E. were supported by the Australian Research Council grant CE170100039. Work at the University of Maryland was supported by NSF award no. DMR-1905891 and the Gordon and Betty Moore Foundation's EPiQS Initiative through grant no. GBMF9071. The work at Los Alamos National Laboratory was performed under the auspices of the U.S. Department of Energy, Office of Basic Energy Sciences, Division of Materials Sciences and Engineering. The work at the University of Warwick was supported by EPSRC, UK through Grant EP/T005963/1. The work at the Johns Hopkins University was supported as part of the Institute for Quantum Matter, an Energy Frontier Research Center funded by the U.S. Department of Energy, Office of Science, Basic Energy Sciences under Award No. DE-SC0019331. Device fabrication was performed in part at the University of Michigan Lurie Nanofabrication Facility. The authors acknowledge the financial support of the University of Michigan College of Engineering and NSF grant \#DMR-0320740, and technical support from the Michigan Center for Materials Characterization.
\end{acknowledgments}

\bibliography{allsmb6papers}

\providecommand{\noopsort}[1]{}\providecommand{\singleletter}[1]{#1}%
\begin{thebibliography}{69}%
\makeatletter
\providecommand \@ifxundefined [1]{%
 \@ifx{#1\undefined}
}%
\providecommand \@ifnum [1]{%
 \ifnum #1\expandafter \@firstoftwo
 \else \expandafter \@secondoftwo
 \fi
}%
\providecommand \@ifx [1]{%
 \ifx #1\expandafter \@firstoftwo
 \else \expandafter \@secondoftwo
 \fi
}%
\providecommand \natexlab [1]{#1}%
\providecommand \enquote  [1]{``#1''}%
\providecommand \bibnamefont  [1]{#1}%
\providecommand \bibfnamefont [1]{#1}%
\providecommand \citenamefont [1]{#1}%
\providecommand \href@noop [0]{\@secondoftwo}%
\providecommand \href [0]{\begingroup \@sanitize@url \@href}%
\providecommand \@href[1]{\@@startlink{#1}\@@href}%
\providecommand \@@href[1]{\endgroup#1\@@endlink}%
\providecommand \@sanitize@url [0]{\catcode `\\12\catcode `\$12\catcode
  `\&12\catcode `\#12\catcode `\^12\catcode `\_12\catcode `\%12\relax}%
\providecommand \@@startlink[1]{}%
\providecommand \@@endlink[0]{}%
\providecommand \url  [0]{\begingroup\@sanitize@url \@url }%
\providecommand \@url [1]{\endgroup\@href {#1}{\urlprefix }}%
\providecommand \urlprefix  [0]{URL }%
\providecommand \Eprint [0]{\href }%
\providecommand \doibase [0]{http://dx.doi.org/}%
\providecommand \selectlanguage [0]{\@gobble}%
\providecommand \bibinfo  [0]{\@secondoftwo}%
\providecommand \bibfield  [0]{\@secondoftwo}%
\providecommand \translation [1]{[#1]}%
\providecommand \BibitemOpen [0]{}%
\providecommand \bibitemStop [0]{}%
\providecommand \bibitemNoStop [0]{.\EOS\space}%
\providecommand \EOS [0]{\spacefactor3000\relax}%
\providecommand \BibitemShut  [1]{\csname bibitem#1\endcsname}%
\let\auto@bib@innerbib\@empty
\bibitem [{\citenamefont {{G. Aeppli and Z. Fisk}}(1992)}]{aeppli}%
  \BibitemOpen
  \bibfield  {author} {\bibinfo {author} {\bibnamefont {{G. Aeppli and Z.
  Fisk}}},\ }\href@noop {} {\bibfield  {journal} {\bibinfo  {journal} {Comments
  Cond. Mat. Phys.}\ }\textbf {\bibinfo {volume} {16}},\ \bibinfo {pages} {155}
  (\bibinfo {year} {1992})}\BibitemShut {NoStop}%
\bibitem [{\citenamefont {Riseborough}(2000)}]{riseborough00}%
  \BibitemOpen
  \bibfield  {author} {\bibinfo {author} {\bibfnamefont {P.~S.}\ \bibnamefont
  {Riseborough}},\ }\href@noop {} {\bibfield  {journal} {\bibinfo  {journal}
  {Advances in Physics}\ }\textbf {\bibinfo {volume} {49}},\ \bibinfo {pages}
  {257} (\bibinfo {year} {2000})}\BibitemShut {NoStop}%
\bibitem [{\citenamefont {Mott}(1974)}]{mott74}%
  \BibitemOpen
  \bibfield  {author} {\bibinfo {author} {\bibfnamefont {N.~F.}\ \bibnamefont
  {Mott}},\ }\href@noop {} {\bibfield  {journal} {\bibinfo  {journal}
  {Philosophical Magazine}\ }\textbf {\bibinfo {volume} {30}},\ \bibinfo
  {pages} {403} (\bibinfo {year} {1974})}\BibitemShut {NoStop}%
\bibitem [{\citenamefont {Menth}\ \emph {et~al.}(1969)\citenamefont {Menth},
  \citenamefont {Buehler},\ and\ \citenamefont {Geballe}}]{menth}%
  \BibitemOpen
  \bibfield  {author} {\bibinfo {author} {\bibfnamefont {A.}~\bibnamefont
  {Menth}}, \bibinfo {author} {\bibfnamefont {E.}~\bibnamefont {Buehler}}, \
  and\ \bibinfo {author} {\bibfnamefont {T.~H.}\ \bibnamefont {Geballe}},\
  }\href@noop {} {\bibfield  {journal} {\bibinfo  {journal} {Physical Review
  Letters}\ }\textbf {\bibinfo {volume} {22}},\ \bibinfo {pages} {295}
  (\bibinfo {year} {1969})}\BibitemShut {NoStop}%
\bibitem [{\citenamefont {Allen}\ \emph {et~al.}(1979)\citenamefont {Allen},
  \citenamefont {Batlogg},\ and\ \citenamefont {Wachter}}]{allen}%
  \BibitemOpen
  \bibfield  {author} {\bibinfo {author} {\bibfnamefont {J.~W.}\ \bibnamefont
  {Allen}}, \bibinfo {author} {\bibfnamefont {B.}~\bibnamefont {Batlogg}}, \
  and\ \bibinfo {author} {\bibfnamefont {P.}~\bibnamefont {Wachter}},\
  }\href@noop {} {\bibfield  {journal} {\bibinfo  {journal} {Physical Review
  B}\ }\textbf {\bibinfo {volume} {20}},\ \bibinfo {pages} {4807} (\bibinfo
  {year} {1979})}\BibitemShut {NoStop}%
\bibitem [{\citenamefont {Martin}\ and\ \citenamefont
  {Allen}(1979)}]{martin1979theory}%
  \BibitemOpen
  \bibfield  {author} {\bibinfo {author} {\bibfnamefont {R.~M.}\ \bibnamefont
  {Martin}}\ and\ \bibinfo {author} {\bibfnamefont {J.}~\bibnamefont {Allen}},\
  }\href@noop {} {\bibfield  {journal} {\bibinfo  {journal} {Journal of Applied
  Physics}\ }\textbf {\bibinfo {volume} {50}},\ \bibinfo {pages} {7561}
  (\bibinfo {year} {1979})}\BibitemShut {NoStop}%
\bibitem [{\citenamefont {Dzero}\ \emph {et~al.}(2010)\citenamefont {Dzero},
  \citenamefont {Sun}, \citenamefont {Galitski},\ and\ \citenamefont
  {Coleman}}]{dzero10}%
  \BibitemOpen
  \bibfield  {author} {\bibinfo {author} {\bibfnamefont {M.}~\bibnamefont
  {Dzero}}, \bibinfo {author} {\bibfnamefont {K.}~\bibnamefont {Sun}}, \bibinfo
  {author} {\bibfnamefont {V.}~\bibnamefont {Galitski}}, \ and\ \bibinfo
  {author} {\bibfnamefont {P.}~\bibnamefont {Coleman}},\ }\href@noop {}
  {\bibfield  {journal} {\bibinfo  {journal} {Physical Review Letters}\
  }\textbf {\bibinfo {volume} {104}},\ \bibinfo {pages} {106408} (\bibinfo
  {year} {2010})}\BibitemShut {NoStop}%
\bibitem [{\citenamefont {Takimoto}(2011)}]{takimoto}%
  \BibitemOpen
  \bibfield  {author} {\bibinfo {author} {\bibfnamefont {T.}~\bibnamefont
  {Takimoto}},\ }\href@noop {} {\bibfield  {journal} {\bibinfo  {journal}
  {Journal of the Physical Society of Japan}\ }\textbf {\bibinfo {volume}
  {80}},\ \bibinfo {pages} {123710} (\bibinfo {year} {2011})}\BibitemShut
  {NoStop}%
\bibitem [{\citenamefont {Dzero}\ \emph {et~al.}(2012)\citenamefont {Dzero},
  \citenamefont {Sun}, \citenamefont {Coleman},\ and\ \citenamefont
  {Galitski}}]{dzero12}%
  \BibitemOpen
  \bibfield  {author} {\bibinfo {author} {\bibfnamefont {M.}~\bibnamefont
  {Dzero}}, \bibinfo {author} {\bibfnamefont {K.}~\bibnamefont {Sun}}, \bibinfo
  {author} {\bibfnamefont {P.}~\bibnamefont {Coleman}}, \ and\ \bibinfo
  {author} {\bibfnamefont {V.}~\bibnamefont {Galitski}},\ }\href@noop {}
  {\bibfield  {journal} {\bibinfo  {journal} {Physical Review B}\ }\textbf
  {\bibinfo {volume} {85}},\ \bibinfo {pages} {045130} (\bibinfo {year}
  {2012})}\BibitemShut {NoStop}%
\bibitem [{\citenamefont {Wolgast}\ \emph {et~al.}(2013)\citenamefont
  {Wolgast}, \citenamefont {\c{C}. Kurdak}, \citenamefont {Sun}, \citenamefont
  {Allen}, \citenamefont {Kim},\ and\ \citenamefont {Fisk}}]{wolgast13}%
  \BibitemOpen
  \bibfield  {author} {\bibinfo {author} {\bibfnamefont {S.}~\bibnamefont
  {Wolgast}}, \bibinfo {author} {\bibnamefont {\c{C}. Kurdak}}, \bibinfo
  {author} {\bibfnamefont {K.}~\bibnamefont {Sun}}, \bibinfo {author}
  {\bibfnamefont {J.~W.}\ \bibnamefont {Allen}}, \bibinfo {author}
  {\bibfnamefont {D.-J.}\ \bibnamefont {Kim}}, \ and\ \bibinfo {author}
  {\bibfnamefont {Z.}~\bibnamefont {Fisk}},\ }\href@noop {} {\bibfield
  {journal} {\bibinfo  {journal} {Physical Review B}\ }\textbf {\bibinfo
  {volume} {88}},\ \bibinfo {pages} {180405(R)} (\bibinfo {year}
  {2013})}\BibitemShut {NoStop}%
\bibitem [{\citenamefont {Kim}\ \emph {et~al.}(2013)\citenamefont {Kim},
  \citenamefont {Thomas}, \citenamefont {Grant}, \citenamefont {Botimer},
  \citenamefont {Fisk},\ and\ \citenamefont {Xia}}]{kim13}%
  \BibitemOpen
  \bibfield  {author} {\bibinfo {author} {\bibfnamefont {D.~J.}\ \bibnamefont
  {Kim}}, \bibinfo {author} {\bibfnamefont {S.}~\bibnamefont {Thomas}},
  \bibinfo {author} {\bibfnamefont {T.}~\bibnamefont {Grant}}, \bibinfo
  {author} {\bibfnamefont {J.}~\bibnamefont {Botimer}}, \bibinfo {author}
  {\bibfnamefont {Z.}~\bibnamefont {Fisk}}, \ and\ \bibinfo {author}
  {\bibfnamefont {J.}~\bibnamefont {Xia}},\ }\href@noop {} {\bibfield
  {journal} {\bibinfo  {journal} {Scientific Reports}\ }\textbf {\bibinfo
  {volume} {3}},\ \bibinfo {pages} {3150} (\bibinfo {year} {2013})}\BibitemShut
  {NoStop}%
\bibitem [{\citenamefont {Kim}\ \emph {et~al.}(2014)\citenamefont {Kim},
  \citenamefont {Xia},\ and\ \citenamefont {Fisk}}]{kim14}%
  \BibitemOpen
  \bibfield  {author} {\bibinfo {author} {\bibfnamefont {D.~J.}\ \bibnamefont
  {Kim}}, \bibinfo {author} {\bibfnamefont {J.}~\bibnamefont {Xia}}, \ and\
  \bibinfo {author} {\bibfnamefont {Z.}~\bibnamefont {Fisk}},\ }\href@noop {}
  {\bibfield  {journal} {\bibinfo  {journal} {Scientific Reports}\ }\textbf
  {\bibinfo {volume} {13}},\ \bibinfo {pages} {466} (\bibinfo {year}
  {2014})}\BibitemShut {NoStop}%
\bibitem [{\citenamefont {Neupane}\ \emph {et~al.}(2013)\citenamefont
  {Neupane}, \citenamefont {Alidoust}, \citenamefont {Xu}, \citenamefont
  {Kondo}, \citenamefont {Ishida}, \citenamefont {Kim}, \citenamefont {Liu},
  \citenamefont {Belopolski}, \citenamefont {Jo}, \citenamefont {Chang},
  \citenamefont {Jeng}, \citenamefont {Durakiewicz}, \citenamefont {Balicas},
  \citenamefont {Lin}, \citenamefont {Bansil}, \citenamefont {Shin},
  \citenamefont {Fisk},\ and\ \citenamefont {Hasan}}]{neupane}%
  \BibitemOpen
  \bibfield  {author} {\bibinfo {author} {\bibfnamefont {M.}~\bibnamefont
  {Neupane}}, \bibinfo {author} {\bibfnamefont {N.}~\bibnamefont {Alidoust}},
  \bibinfo {author} {\bibfnamefont {S.-Y.}\ \bibnamefont {Xu}}, \bibinfo
  {author} {\bibfnamefont {T.}~\bibnamefont {Kondo}}, \bibinfo {author}
  {\bibfnamefont {Y.}~\bibnamefont {Ishida}}, \bibinfo {author} {\bibfnamefont
  {D.~J.}\ \bibnamefont {Kim}}, \bibinfo {author} {\bibfnamefont
  {C.}~\bibnamefont {Liu}}, \bibinfo {author} {\bibfnamefont {I.}~\bibnamefont
  {Belopolski}}, \bibinfo {author} {\bibfnamefont {Y.~J.}\ \bibnamefont {Jo}},
  \bibinfo {author} {\bibfnamefont {T.-R.}\ \bibnamefont {Chang}}, \bibinfo
  {author} {\bibfnamefont {H.-T.}\ \bibnamefont {Jeng}}, \bibinfo {author}
  {\bibfnamefont {T.}~\bibnamefont {Durakiewicz}}, \bibinfo {author}
  {\bibfnamefont {L.}~\bibnamefont {Balicas}}, \bibinfo {author} {\bibfnamefont
  {H.}~\bibnamefont {Lin}}, \bibinfo {author} {\bibfnamefont {A.}~\bibnamefont
  {Bansil}}, \bibinfo {author} {\bibfnamefont {S.}~\bibnamefont {Shin}},
  \bibinfo {author} {\bibfnamefont {Z.}~\bibnamefont {Fisk}}, \ and\ \bibinfo
  {author} {\bibfnamefont {M.~Z.}\ \bibnamefont {Hasan}},\ }\href@noop {}
  {\bibfield  {journal} {\bibinfo  {journal} {Nature Communications}\ }\textbf
  {\bibinfo {volume} {4}},\ \bibinfo {pages} {2991} (\bibinfo {year}
  {2013})}\BibitemShut {NoStop}%
\bibitem [{\citenamefont {Jiang}\ \emph {et~al.}(2013)\citenamefont {Jiang},
  \citenamefont {Li}, \citenamefont {Zhang}, \citenamefont {Sun}, \citenamefont
  {Chen}, \citenamefont {Ye}, \citenamefont {Xu}, \citenamefont {Ge},
  \citenamefont {Tan}, \citenamefont {Niu}, \citenamefont {Xia}, \citenamefont
  {Xie}, \citenamefont {Li}, \citenamefont {Chen}, \citenamefont {Wen},\ and\
  \citenamefont {Feng}}]{jiang}%
  \BibitemOpen
  \bibfield  {author} {\bibinfo {author} {\bibfnamefont {J.}~\bibnamefont
  {Jiang}}, \bibinfo {author} {\bibfnamefont {S.}~\bibnamefont {Li}}, \bibinfo
  {author} {\bibfnamefont {T.}~\bibnamefont {Zhang}}, \bibinfo {author}
  {\bibfnamefont {Z.}~\bibnamefont {Sun}}, \bibinfo {author} {\bibfnamefont
  {F.}~\bibnamefont {Chen}}, \bibinfo {author} {\bibfnamefont {Z.~R.}\
  \bibnamefont {Ye}}, \bibinfo {author} {\bibfnamefont {M.}~\bibnamefont {Xu}},
  \bibinfo {author} {\bibfnamefont {Q.~Q.}\ \bibnamefont {Ge}}, \bibinfo
  {author} {\bibfnamefont {S.~Y.}\ \bibnamefont {Tan}}, \bibinfo {author}
  {\bibfnamefont {X.~H.}\ \bibnamefont {Niu}}, \bibinfo {author} {\bibfnamefont
  {M.}~\bibnamefont {Xia}}, \bibinfo {author} {\bibfnamefont {B.~P.}\
  \bibnamefont {Xie}}, \bibinfo {author} {\bibfnamefont {Y.~F.}\ \bibnamefont
  {Li}}, \bibinfo {author} {\bibfnamefont {X.~H.}\ \bibnamefont {Chen}},
  \bibinfo {author} {\bibfnamefont {H.~H.}\ \bibnamefont {Wen}}, \ and\
  \bibinfo {author} {\bibfnamefont {D.~L.}\ \bibnamefont {Feng}},\ }\href@noop
  {} {\bibfield  {journal} {\bibinfo  {journal} {Nature Communications}\
  }\textbf {\bibinfo {volume} {4}},\ \bibinfo {pages} {3010} (\bibinfo {year}
  {2013})}\BibitemShut {NoStop}%
\bibitem [{\citenamefont {Xu}\ \emph {et~al.}(2014)\citenamefont {Xu},
  \citenamefont {Biswas}, \citenamefont {Dil}, \citenamefont {Dhaka},
  \citenamefont {Landolt}, \citenamefont {Muff}, \citenamefont {Matt},
  \citenamefont {Shi}, \citenamefont {Plumb}, \citenamefont {Radovic},
  \citenamefont {Pomjakushina}, \citenamefont {Conder}, \citenamefont {Amato},
  \citenamefont {Borisenko}, \citenamefont {Yu}, \citenamefont {Weng},
  \citenamefont {Fang}, \citenamefont {Dai}, \citenamefont {Mesot},
  \citenamefont {Ding},\ and\ \citenamefont {Shi}}]{nxu14}%
  \BibitemOpen
  \bibfield  {author} {\bibinfo {author} {\bibfnamefont {N.}~\bibnamefont
  {Xu}}, \bibinfo {author} {\bibfnamefont {P.~K.}\ \bibnamefont {Biswas}},
  \bibinfo {author} {\bibfnamefont {J.~H.}\ \bibnamefont {Dil}}, \bibinfo
  {author} {\bibfnamefont {R.~S.}\ \bibnamefont {Dhaka}}, \bibinfo {author}
  {\bibfnamefont {G.}~\bibnamefont {Landolt}}, \bibinfo {author} {\bibfnamefont
  {S.}~\bibnamefont {Muff}}, \bibinfo {author} {\bibfnamefont {C.~E.}\
  \bibnamefont {Matt}}, \bibinfo {author} {\bibfnamefont {X.}~\bibnamefont
  {Shi}}, \bibinfo {author} {\bibfnamefont {N.~C.}\ \bibnamefont {Plumb}},
  \bibinfo {author} {\bibfnamefont {M.}~\bibnamefont {Radovic}}, \bibinfo
  {author} {\bibfnamefont {E.}~\bibnamefont {Pomjakushina}}, \bibinfo {author}
  {\bibfnamefont {K.}~\bibnamefont {Conder}}, \bibinfo {author} {\bibfnamefont
  {A.}~\bibnamefont {Amato}}, \bibinfo {author} {\bibfnamefont {S.~V.}\
  \bibnamefont {Borisenko}}, \bibinfo {author} {\bibfnamefont {R.}~\bibnamefont
  {Yu}}, \bibinfo {author} {\bibfnamefont {H.~M.}\ \bibnamefont {Weng}},
  \bibinfo {author} {\bibfnamefont {Z.}~\bibnamefont {Fang}}, \bibinfo {author}
  {\bibfnamefont {X.}~\bibnamefont {Dai}}, \bibinfo {author} {\bibfnamefont
  {J.}~\bibnamefont {Mesot}}, \bibinfo {author} {\bibfnamefont
  {H.}~\bibnamefont {Ding}}, \ and\ \bibinfo {author} {\bibfnamefont
  {M.}~\bibnamefont {Shi}},\ }\href@noop {} {\bibfield  {journal} {\bibinfo
  {journal} {Nature Communications}\ }\textbf {\bibinfo {volume} {5}},\
  \bibinfo {pages} {4566} (\bibinfo {year} {2014})}\BibitemShut {NoStop}%
\bibitem [{\citenamefont {Denlinger}\ \emph {et~al.}(2014)\citenamefont
  {Denlinger}, \citenamefont {Allen}, \citenamefont {Kang}, \citenamefont
  {Sun}, \citenamefont {Kim}, \citenamefont {Shim}, \citenamefont {Min},
  \citenamefont {Kim},\ and\ \citenamefont {Fisk}}]{denlingerlinked}%
  \BibitemOpen
  \bibfield  {author} {\bibinfo {author} {\bibfnamefont {J.~D.}\ \bibnamefont
  {Denlinger}}, \bibinfo {author} {\bibfnamefont {J.~W.}\ \bibnamefont
  {Allen}}, \bibinfo {author} {\bibfnamefont {J.-S.}\ \bibnamefont {Kang}},
  \bibinfo {author} {\bibfnamefont {K.}~\bibnamefont {Sun}}, \bibinfo {author}
  {\bibfnamefont {J.-W.}\ \bibnamefont {Kim}}, \bibinfo {author} {\bibfnamefont
  {J.~H.}\ \bibnamefont {Shim}}, \bibinfo {author} {\bibfnamefont {B.~I.}\
  \bibnamefont {Min}}, \bibinfo {author} {\bibfnamefont {D.-J.}\ \bibnamefont
  {Kim}}, \ and\ \bibinfo {author} {\bibfnamefont {Z.}~\bibnamefont {Fisk}},\
  }\href@noop {} {\enquote {\bibinfo {title} {{Temperature dependence of linked
  gap and surface state evolution in the mixed valent topological insulator
  {S}m{B}$_{6}$}},}\ } (\bibinfo {year} {2014}),\ \bibinfo {note}
  {{arXiv:1312.6637 [cond-mat.str-el]}}\BibitemShut {NoStop}%
\bibitem [{\citenamefont {Zhang}\ \emph {et~al.}(2013)\citenamefont {Zhang},
  \citenamefont {Butch}, \citenamefont {Syers}, \citenamefont {Ziemak},
  \citenamefont {Greene},\ and\ \citenamefont {Paglione}}]{zhang13}%
  \BibitemOpen
  \bibfield  {author} {\bibinfo {author} {\bibfnamefont {X.}~\bibnamefont
  {Zhang}}, \bibinfo {author} {\bibfnamefont {N.~P.}\ \bibnamefont {Butch}},
  \bibinfo {author} {\bibfnamefont {P.}~\bibnamefont {Syers}}, \bibinfo
  {author} {\bibfnamefont {S.}~\bibnamefont {Ziemak}}, \bibinfo {author}
  {\bibfnamefont {R.~L.}\ \bibnamefont {Greene}}, \ and\ \bibinfo {author}
  {\bibfnamefont {J.}~\bibnamefont {Paglione}},\ }\href@noop {} {\bibfield
  {journal} {\bibinfo  {journal} {Physical Review X}\ }\textbf {\bibinfo
  {volume} {3}},\ \bibinfo {pages} {011011} (\bibinfo {year}
  {2013})}\BibitemShut {NoStop}%
\bibitem [{\citenamefont {R{\"o}{\ss}ler}\ \emph {et~al.}(2016)\citenamefont
  {R{\"o}{\ss}ler}, \citenamefont {Jiao}, \citenamefont {Kim}, \citenamefont
  {Seiro}, \citenamefont {Rasim}, \citenamefont {Steglich}, \citenamefont
  {Tjeng}, \citenamefont {Fisk},\ and\ \citenamefont {Wirth}}]{rossler16}%
  \BibitemOpen
  \bibfield  {author} {\bibinfo {author} {\bibfnamefont {S.}~\bibnamefont
  {R{\"o}{\ss}ler}}, \bibinfo {author} {\bibfnamefont {L.}~\bibnamefont
  {Jiao}}, \bibinfo {author} {\bibfnamefont {D.-J.}\ \bibnamefont {Kim}},
  \bibinfo {author} {\bibfnamefont {S.}~\bibnamefont {Seiro}}, \bibinfo
  {author} {\bibfnamefont {K.}~\bibnamefont {Rasim}}, \bibinfo {author}
  {\bibfnamefont {F.}~\bibnamefont {Steglich}}, \bibinfo {author}
  {\bibfnamefont {L.~H.}\ \bibnamefont {Tjeng}}, \bibinfo {author}
  {\bibfnamefont {Z.}~\bibnamefont {Fisk}}, \ and\ \bibinfo {author}
  {\bibfnamefont {S.}~\bibnamefont {Wirth}},\ }\href@noop {} {\bibfield
  {journal} {\bibinfo  {journal} {Philosophical Magazine}\ }\textbf {\bibinfo
  {volume} {96}},\ \bibinfo {pages} {3262} (\bibinfo {year}
  {2016})}\BibitemShut {NoStop}%
\bibitem [{\citenamefont {Pirie}\ \emph {et~al.}(2020)\citenamefont {Pirie},
  \citenamefont {Liu}, \citenamefont {Soumyanarayanan}, \citenamefont {Chen},
  \citenamefont {He}, \citenamefont {Yee}, \citenamefont {Rosa}, \citenamefont
  {Thompson}, \citenamefont {Kim}, \citenamefont {Fisk}, \citenamefont {Wang},
  \citenamefont {Paglione}, \citenamefont {Morr}, \citenamefont {Hamidian},\
  and\ \citenamefont {Hoffmann}}]{pirie}%
  \BibitemOpen
  \bibfield  {author} {\bibinfo {author} {\bibfnamefont {H.}~\bibnamefont
  {Pirie}}, \bibinfo {author} {\bibfnamefont {Y.}~\bibnamefont {Liu}}, \bibinfo
  {author} {\bibfnamefont {A.}~\bibnamefont {Soumyanarayanan}}, \bibinfo
  {author} {\bibfnamefont {P.}~\bibnamefont {Chen}}, \bibinfo {author}
  {\bibfnamefont {Y.}~\bibnamefont {He}}, \bibinfo {author} {\bibfnamefont
  {M.~M.}\ \bibnamefont {Yee}}, \bibinfo {author} {\bibfnamefont {P.~F.~S.}\
  \bibnamefont {Rosa}}, \bibinfo {author} {\bibfnamefont {J.~D.}\ \bibnamefont
  {Thompson}}, \bibinfo {author} {\bibfnamefont {D.-J.}\ \bibnamefont {Kim}},
  \bibinfo {author} {\bibfnamefont {Z.}~\bibnamefont {Fisk}}, \bibinfo {author}
  {\bibfnamefont {X.}~\bibnamefont {Wang}}, \bibinfo {author} {\bibfnamefont
  {J.}~\bibnamefont {Paglione}}, \bibinfo {author} {\bibfnamefont {D.~K.}\
  \bibnamefont {Morr}}, \bibinfo {author} {\bibfnamefont {M.~H.}\ \bibnamefont
  {Hamidian}}, \ and\ \bibinfo {author} {\bibfnamefont {J.~E.}\ \bibnamefont
  {Hoffmann}},\ }\href@noop {} {\bibfield  {journal} {\bibinfo  {journal}
  {Nature Physics}\ }\textbf {\bibinfo {volume} {16}},\ \bibinfo {pages} {52}
  (\bibinfo {year} {2020})}\BibitemShut {NoStop}%
\bibitem [{\citenamefont {Fuhrman}\ \emph {et~al.}(2015)\citenamefont
  {Fuhrman}, \citenamefont {Leiner}, \citenamefont {Nikoli\'{c}}, \citenamefont
  {Granroth}, \citenamefont {Stone}, \citenamefont {Lumsden}, \citenamefont
  {{DeBeer-Schmidtt}}, \citenamefont {Alekseev}, \citenamefont {Mignot},
  \citenamefont {Mignot}, \citenamefont {Koohpayeh}, \citenamefont
  {Cottingham}, \citenamefont {Phelan}, \citenamefont {Schoop}, \citenamefont
  {Schoop}, \citenamefont {McQueen},\ and\ \citenamefont
  {Broholm}}]{fuhrman15}%
  \BibitemOpen
  \bibfield  {author} {\bibinfo {author} {\bibfnamefont {W.~T.}\ \bibnamefont
  {Fuhrman}}, \bibinfo {author} {\bibfnamefont {J.}~\bibnamefont {Leiner}},
  \bibinfo {author} {\bibfnamefont {P.}~\bibnamefont {Nikoli\'{c}}}, \bibinfo
  {author} {\bibfnamefont {G.~E.}\ \bibnamefont {Granroth}}, \bibinfo {author}
  {\bibfnamefont {M.~B.}\ \bibnamefont {Stone}}, \bibinfo {author}
  {\bibfnamefont {M.~D.}\ \bibnamefont {Lumsden}}, \bibinfo {author}
  {\bibfnamefont {L.}~\bibnamefont {{DeBeer-Schmidtt}}}, \bibinfo {author}
  {\bibfnamefont {P.~A.}\ \bibnamefont {Alekseev}}, \bibinfo {author}
  {\bibfnamefont {J.-M.}\ \bibnamefont {Mignot}}, \bibinfo {author}
  {\bibfnamefont {S.~M.}\ \bibnamefont {Mignot}}, \bibinfo {author}
  {\bibfnamefont {S.~M.}\ \bibnamefont {Koohpayeh}}, \bibinfo {author}
  {\bibfnamefont {P.}~\bibnamefont {Cottingham}}, \bibinfo {author}
  {\bibfnamefont {W.~A.}\ \bibnamefont {Phelan}}, \bibinfo {author}
  {\bibfnamefont {L.}~\bibnamefont {Schoop}}, \bibinfo {author} {\bibfnamefont
  {T.~M.}\ \bibnamefont {Schoop}}, \bibinfo {author} {\bibfnamefont {T.~M.}\
  \bibnamefont {McQueen}}, \ and\ \bibinfo {author} {\bibfnamefont
  {C.}~\bibnamefont {Broholm}},\ }\href@noop {} {\bibfield  {journal} {\bibinfo
   {journal} {Physical Review Letters}\ }\textbf {\bibinfo {volume} {114}},\
  \bibinfo {pages} {036401} (\bibinfo {year} {2015})}\BibitemShut {NoStop}%
\bibitem [{\citenamefont {Hlawenka}\ \emph {et~al.}(2018)\citenamefont
  {Hlawenka}, \citenamefont {Siemensmeyer}, \citenamefont {Weschke},
  \citenamefont {Varykhalov}, \citenamefont {S\'{a}nchez-Barriga},
  \citenamefont {Shitsevalova}, \citenamefont {Dukhnenko}, \citenamefont
  {Filipov}, \citenamefont {Gab\'{a}ni}, \citenamefont {Flachbart},
  \citenamefont {Rader},\ and\ \citenamefont {Rienks}}]{hlawenka}%
  \BibitemOpen
  \bibfield  {author} {\bibinfo {author} {\bibfnamefont {P.}~\bibnamefont
  {Hlawenka}}, \bibinfo {author} {\bibfnamefont {K.}~\bibnamefont
  {Siemensmeyer}}, \bibinfo {author} {\bibfnamefont {E.}~\bibnamefont
  {Weschke}}, \bibinfo {author} {\bibfnamefont {A.}~\bibnamefont {Varykhalov}},
  \bibinfo {author} {\bibfnamefont {J.}~\bibnamefont {S\'{a}nchez-Barriga}},
  \bibinfo {author} {\bibfnamefont {N.~Y.}\ \bibnamefont {Shitsevalova}},
  \bibinfo {author} {\bibfnamefont {A.~V.}\ \bibnamefont {Dukhnenko}}, \bibinfo
  {author} {\bibfnamefont {V.~B.}\ \bibnamefont {Filipov}}, \bibinfo {author}
  {\bibfnamefont {S.}~\bibnamefont {Gab\'{a}ni}}, \bibinfo {author}
  {\bibfnamefont {K.}~\bibnamefont {Flachbart}}, \bibinfo {author}
  {\bibfnamefont {O.}~\bibnamefont {Rader}}, \ and\ \bibinfo {author}
  {\bibfnamefont {E.~D.~L.}\ \bibnamefont {Rienks}},\ }\href@noop {} {\bibfield
   {journal} {\bibinfo  {journal} {Nature Communications}\ }\textbf {\bibinfo
  {volume} {9}},\ \bibinfo {pages} {517} (\bibinfo {year} {2018})}\BibitemShut
  {NoStop}%
\bibitem [{\citenamefont {Hermann}\ \emph {et~al.}(2020)\citenamefont
  {Hermann}, \citenamefont {Hlawenka}, \citenamefont {Siemensmeier},
  \citenamefont {Weschke}, \citenamefont {S\'{a}nchez-Barriga}, \citenamefont
  {Varykhalov}, \citenamefont {Shitsevalova}, \citenamefont {Dukhnenko},
  \citenamefont {Filipov}, \citenamefont {Gab\'{a}ni}, \citenamefont
  {Flachbart}, \citenamefont {Rader}, \citenamefont {Sterrer},\ and\
  \citenamefont {Rienks}}]{hermann}%
  \BibitemOpen
  \bibfield  {author} {\bibinfo {author} {\bibfnamefont {H.}~\bibnamefont
  {Hermann}}, \bibinfo {author} {\bibfnamefont {P.}~\bibnamefont {Hlawenka}},
  \bibinfo {author} {\bibfnamefont {K.}~\bibnamefont {Siemensmeier}}, \bibinfo
  {author} {\bibfnamefont {E.}~\bibnamefont {Weschke}}, \bibinfo {author}
  {\bibfnamefont {J.}~\bibnamefont {S\'{a}nchez-Barriga}}, \bibinfo {author}
  {\bibfnamefont {A.}~\bibnamefont {Varykhalov}}, \bibinfo {author}
  {\bibfnamefont {N.~Y.}\ \bibnamefont {Shitsevalova}}, \bibinfo {author}
  {\bibfnamefont {A.~V.}\ \bibnamefont {Dukhnenko}}, \bibinfo {author}
  {\bibfnamefont {V.~B.}\ \bibnamefont {Filipov}}, \bibinfo {author}
  {\bibfnamefont {S.}~\bibnamefont {Gab\'{a}ni}}, \bibinfo {author}
  {\bibfnamefont {K.}~\bibnamefont {Flachbart}}, \bibinfo {author}
  {\bibfnamefont {O.}~\bibnamefont {Rader}}, \bibinfo {author} {\bibfnamefont
  {M.}~\bibnamefont {Sterrer}}, \ and\ \bibinfo {author} {\bibfnamefont
  {E.~D.~L.}\ \bibnamefont {Rienks}},\ }\href@noop {} {\bibfield  {journal}
  {\bibinfo  {journal} {Advanced Materials}\ }\textbf {\bibinfo {volume}
  {32}},\ \bibinfo {pages} {1906725} (\bibinfo {year} {2020})}\BibitemShut
  {NoStop}%
\bibitem [{\citenamefont {Frantzeskakis}\ \emph {et~al.}(2013)\citenamefont
  {Frantzeskakis}, \citenamefont {{de Jong}}, \citenamefont {Zwartsenberg},
  \citenamefont {Huang}, \citenamefont {Pan}, \citenamefont {Zhang},
  \citenamefont {Zhang}, \citenamefont {Zhang}, \citenamefont {Bao},
  \citenamefont {Tegus}, \citenamefont {Varykhalov}, \citenamefont {{de
  Visser}},\ and\ \citenamefont {Golden}}]{frantzeskakis}%
  \BibitemOpen
  \bibfield  {author} {\bibinfo {author} {\bibfnamefont {E.}~\bibnamefont
  {Frantzeskakis}}, \bibinfo {author} {\bibfnamefont {N.}~\bibnamefont {{de
  Jong}}}, \bibinfo {author} {\bibfnamefont {B.}~\bibnamefont {Zwartsenberg}},
  \bibinfo {author} {\bibfnamefont {Y.~K.}\ \bibnamefont {Huang}}, \bibinfo
  {author} {\bibfnamefont {Y.}~\bibnamefont {Pan}}, \bibinfo {author}
  {\bibfnamefont {X.}~\bibnamefont {Zhang}}, \bibinfo {author} {\bibfnamefont
  {J.~X.}\ \bibnamefont {Zhang}}, \bibinfo {author} {\bibfnamefont {F.~X.}\
  \bibnamefont {Zhang}}, \bibinfo {author} {\bibfnamefont {L.~H.}\ \bibnamefont
  {Bao}}, \bibinfo {author} {\bibfnamefont {O.}~\bibnamefont {Tegus}}, \bibinfo
  {author} {\bibfnamefont {A.}~\bibnamefont {Varykhalov}}, \bibinfo {author}
  {\bibfnamefont {A.}~\bibnamefont {{de Visser}}}, \ and\ \bibinfo {author}
  {\bibfnamefont {M.~S.}\ \bibnamefont {Golden}},\ }\href@noop {} {\bibfield
  {journal} {\bibinfo  {journal} {Physical Review X}\ }\textbf {\bibinfo
  {volume} {3}},\ \bibinfo {pages} {041024} (\bibinfo {year}
  {2013})}\BibitemShut {NoStop}%
\bibitem [{\citenamefont {Phelan}\ \emph {et~al.}(2016)\citenamefont {Phelan},
  \citenamefont {Koohpayeh}, \citenamefont {Cottingham}, \citenamefont
  {Tutmaher}, \citenamefont {Leiner}, \citenamefont {Lumsden}, \citenamefont
  {Lavelle}, \citenamefont {Wang}, \citenamefont {Hoffmann}, \citenamefont
  {Siegler}, \citenamefont {Haldolaarachchige}, \citenamefont {Young},\ and\
  \citenamefont {McQueen}}]{phelan16}%
  \BibitemOpen
  \bibfield  {author} {\bibinfo {author} {\bibfnamefont {W.~A.}\ \bibnamefont
  {Phelan}}, \bibinfo {author} {\bibfnamefont {S.~M.}\ \bibnamefont
  {Koohpayeh}}, \bibinfo {author} {\bibfnamefont {P.}~\bibnamefont
  {Cottingham}}, \bibinfo {author} {\bibfnamefont {J.~A.}\ \bibnamefont
  {Tutmaher}}, \bibinfo {author} {\bibfnamefont {J.~C.}\ \bibnamefont
  {Leiner}}, \bibinfo {author} {\bibfnamefont {M.~D.}\ \bibnamefont {Lumsden}},
  \bibinfo {author} {\bibfnamefont {C.~M.}\ \bibnamefont {Lavelle}}, \bibinfo
  {author} {\bibfnamefont {X.~P.}\ \bibnamefont {Wang}}, \bibinfo {author}
  {\bibfnamefont {C.}~\bibnamefont {Hoffmann}}, \bibinfo {author}
  {\bibfnamefont {M.~A.}\ \bibnamefont {Siegler}}, \bibinfo {author}
  {\bibfnamefont {N.}~\bibnamefont {Haldolaarachchige}}, \bibinfo {author}
  {\bibfnamefont {D.~P.}\ \bibnamefont {Young}}, \ and\ \bibinfo {author}
  {\bibfnamefont {T.~M.}\ \bibnamefont {McQueen}},\ }\href@noop {} {\bibfield
  {journal} {\bibinfo  {journal} {Scientific Reports}\ }\textbf {\bibinfo
  {volume} {6}},\ \bibinfo {pages} {20860} (\bibinfo {year}
  {2016})}\BibitemShut {NoStop}%
\bibitem [{\citenamefont {{P. C. Canfield and Z. Fisk}}(1992)}]{canfield}%
  \BibitemOpen
  \bibfield  {author} {\bibinfo {author} {\bibnamefont {{P. C. Canfield and Z.
  Fisk}}},\ }\href@noop {} {\bibfield  {journal} {\bibinfo  {journal}
  {Philosophical Magazine B}\ }\textbf {\bibinfo {volume} {{65:6}}},\ \bibinfo
  {pages} {1117} (\bibinfo {year} {1992})}\BibitemShut {NoStop}%
\bibitem [{\citenamefont {Koohpayeh}\ \emph {et~al.}(2008)\citenamefont
  {Koohpayeh}, \citenamefont {Fort},\ and\ \citenamefont {Abell}}]{koohpayeh}%
  \BibitemOpen
  \bibfield  {author} {\bibinfo {author} {\bibfnamefont {S.~M.}\ \bibnamefont
  {Koohpayeh}}, \bibinfo {author} {\bibfnamefont {D.}~\bibnamefont {Fort}}, \
  and\ \bibinfo {author} {\bibfnamefont {J.~S.}\ \bibnamefont {Abell}},\
  }\href@noop {} {\bibfield  {journal} {\bibinfo  {journal} {Nature Chemistry}\
  }\textbf {\bibinfo {volume} {54}},\ \bibinfo {pages} {121} (\bibinfo {year}
  {2008})}\BibitemShut {NoStop}%
\bibitem [{\citenamefont {Hatnean}\ \emph {et~al.}(2013)\citenamefont
  {Hatnean}, \citenamefont {Lees}, \citenamefont {Paul},\ and\ \citenamefont
  {Balakrishnan}}]{hatnean}%
  \BibitemOpen
  \bibfield  {author} {\bibinfo {author} {\bibfnamefont {M.~C.}\ \bibnamefont
  {Hatnean}}, \bibinfo {author} {\bibfnamefont {M.~R.}\ \bibnamefont {Lees}},
  \bibinfo {author} {\bibfnamefont {D.~M.}\ \bibnamefont {Paul}}, \ and\
  \bibinfo {author} {\bibfnamefont {G.}~\bibnamefont {Balakrishnan}},\
  }\href@noop {} {\bibfield  {journal} {\bibinfo  {journal} {Scientific
  Reports}\ }\textbf {\bibinfo {volume} {3}},\ \bibinfo {pages} {307} (\bibinfo
  {year} {2013})}\BibitemShut {NoStop}%
\bibitem [{\citenamefont {Li}\ \emph {et~al.}(2014)\citenamefont {Li},
  \citenamefont {Xiang}, \citenamefont {Yu}, \citenamefont {Asaba},
  \citenamefont {Lawson}, \citenamefont {Cai}, \citenamefont {Tinsman},
  \citenamefont {Berkley}, \citenamefont {Wolgast}, \citenamefont {Eo},
  \citenamefont {Kim}, \citenamefont {Kurdak}, \citenamefont {Allen},
  \citenamefont {Sun}, \citenamefont {Cheng}, \citenamefont {Wang},
  \citenamefont {Fisk},\ and\ \citenamefont {Li}}]{liscience}%
  \BibitemOpen
  \bibfield  {author} {\bibinfo {author} {\bibfnamefont {G.}~\bibnamefont
  {Li}}, \bibinfo {author} {\bibfnamefont {Z.}~\bibnamefont {Xiang}}, \bibinfo
  {author} {\bibfnamefont {F.}~\bibnamefont {Yu}}, \bibinfo {author}
  {\bibfnamefont {T.}~\bibnamefont {Asaba}}, \bibinfo {author} {\bibfnamefont
  {B.}~\bibnamefont {Lawson}}, \bibinfo {author} {\bibfnamefont
  {P.}~\bibnamefont {Cai}}, \bibinfo {author} {\bibfnamefont {C.}~\bibnamefont
  {Tinsman}}, \bibinfo {author} {\bibfnamefont {A.}~\bibnamefont {Berkley}},
  \bibinfo {author} {\bibfnamefont {S.}~\bibnamefont {Wolgast}}, \bibinfo
  {author} {\bibfnamefont {Y.~S.}\ \bibnamefont {Eo}}, \bibinfo {author}
  {\bibfnamefont {D.-J.}\ \bibnamefont {Kim}}, \bibinfo {author} {\bibfnamefont
  {C.}~\bibnamefont {Kurdak}}, \bibinfo {author} {\bibfnamefont {J.~W.}\
  \bibnamefont {Allen}}, \bibinfo {author} {\bibfnamefont {K.}~\bibnamefont
  {Sun}}, \bibinfo {author} {\bibfnamefont {X.~H.}\ \bibnamefont {Cheng}},
  \bibinfo {author} {\bibfnamefont {Y.~Y.}\ \bibnamefont {Wang}}, \bibinfo
  {author} {\bibfnamefont {Z.}~\bibnamefont {Fisk}}, \ and\ \bibinfo {author}
  {\bibfnamefont {L.}~\bibnamefont {Li}},\ }\href@noop {} {\bibfield  {journal}
  {\bibinfo  {journal} {Science}\ }\textbf {\bibinfo {volume} {346}},\ \bibinfo
  {pages} {1208} (\bibinfo {year} {2014})}\BibitemShut {NoStop}%
\bibitem [{\citenamefont {Tan}\ \emph {et~al.}(2015)\citenamefont {Tan},
  \citenamefont {Hsu}, \citenamefont {Zeng}, \citenamefont {{Ciomaga Hatnean}},
  \citenamefont {Harrison}, \citenamefont {Zhu}, \citenamefont {Hartstein},
  \citenamefont {Kiourlappou}, \citenamefont {Srivastava}, \citenamefont
  {Johannes}, \citenamefont {Murphy}, \citenamefont {Park}, \citenamefont
  {Balicas}, \citenamefont {Lonzarich}, \citenamefont {Balakrishnan},\ and\
  \citenamefont {Sebastian}}]{tan}%
  \BibitemOpen
  \bibfield  {author} {\bibinfo {author} {\bibfnamefont {B.~S.}\ \bibnamefont
  {Tan}}, \bibinfo {author} {\bibfnamefont {Y.-T.}\ \bibnamefont {Hsu}},
  \bibinfo {author} {\bibfnamefont {B.}~\bibnamefont {Zeng}}, \bibinfo {author}
  {\bibfnamefont {M.}~\bibnamefont {{Ciomaga Hatnean}}}, \bibinfo {author}
  {\bibfnamefont {N.}~\bibnamefont {Harrison}}, \bibinfo {author}
  {\bibfnamefont {Z.}~\bibnamefont {Zhu}}, \bibinfo {author} {\bibfnamefont
  {M.}~\bibnamefont {Hartstein}}, \bibinfo {author} {\bibfnamefont
  {M.}~\bibnamefont {Kiourlappou}}, \bibinfo {author} {\bibfnamefont
  {A.}~\bibnamefont {Srivastava}}, \bibinfo {author} {\bibfnamefont {M.~D.}\
  \bibnamefont {Johannes}}, \bibinfo {author} {\bibfnamefont {T.~P.}\
  \bibnamefont {Murphy}}, \bibinfo {author} {\bibfnamefont {J.-H.}\
  \bibnamefont {Park}}, \bibinfo {author} {\bibfnamefont {L.}~\bibnamefont
  {Balicas}}, \bibinfo {author} {\bibfnamefont {G.~G.}\ \bibnamefont
  {Lonzarich}}, \bibinfo {author} {\bibfnamefont {G.}~\bibnamefont
  {Balakrishnan}}, \ and\ \bibinfo {author} {\bibfnamefont {S.~E.}\
  \bibnamefont {Sebastian}},\ }\href@noop {} {\bibfield  {journal} {\bibinfo
  {journal} {Science}\ }\textbf {\bibinfo {volume} {349}},\ \bibinfo {pages}
  {6245} (\bibinfo {year} {2015})}\BibitemShut {NoStop}%
\bibitem [{\citenamefont {Xiang}\ \emph {et~al.}(2017)\citenamefont {Xiang},
  \citenamefont {Lawson}, \citenamefont {Asaba}, \citenamefont {Tinsman},
  \citenamefont {Chen}, \citenamefont {Shang}, \citenamefont {Chen},\ and\
  \citenamefont {Li}}]{xiang17}%
  \BibitemOpen
  \bibfield  {author} {\bibinfo {author} {\bibfnamefont {Z.}~\bibnamefont
  {Xiang}}, \bibinfo {author} {\bibfnamefont {B.}~\bibnamefont {Lawson}},
  \bibinfo {author} {\bibfnamefont {T.}~\bibnamefont {Asaba}}, \bibinfo
  {author} {\bibfnamefont {C.}~\bibnamefont {Tinsman}}, \bibinfo {author}
  {\bibfnamefont {L.}~\bibnamefont {Chen}}, \bibinfo {author} {\bibfnamefont
  {C.}~\bibnamefont {Shang}}, \bibinfo {author} {\bibfnamefont {X.~H.}\
  \bibnamefont {Chen}}, \ and\ \bibinfo {author} {\bibfnamefont
  {L.}~\bibnamefont {Li}},\ }\href@noop {} {\bibfield  {journal} {\bibinfo
  {journal} {Physical Review X}\ }\textbf {\bibinfo {volume} {7}},\ \bibinfo
  {pages} {031054} (\bibinfo {year} {2017})}\BibitemShut {NoStop}%
\bibitem [{\citenamefont {Hartstein}\ \emph {et~al.}(2017)\citenamefont
  {Hartstein}, \citenamefont {Toews}, \citenamefont {Hsu}, \citenamefont
  {B.Zeng}, \citenamefont {Chen}, \citenamefont {Hatnean}, \citenamefont
  {Zhang}, \citenamefont {Nakamura}, \citenamefont {Padgett}, \citenamefont
  {Rodway-Gant}, \citenamefont {Berk}, \citenamefont {Kingston}, \citenamefont
  {Zhang}, \citenamefont {Chan}, \citenamefont {Yamashita}, \citenamefont
  {Sakakibara}, \citenamefont {Tanako}, \citenamefont {Park}, \citenamefont
  {Balicas}, \citenamefont {Harrison}, \citenamefont {Shitsevalova},
  \citenamefont {Balakrishnan}, \citenamefont {Lonzarich}, \citenamefont
  {Hill}, \citenamefont {Sutherland},\ and\ \citenamefont
  {Sebastian}}]{hartstein}%
  \BibitemOpen
  \bibfield  {author} {\bibinfo {author} {\bibfnamefont {M.}~\bibnamefont
  {Hartstein}}, \bibinfo {author} {\bibfnamefont {W.~H.}\ \bibnamefont
  {Toews}}, \bibinfo {author} {\bibfnamefont {Y.-T.}\ \bibnamefont {Hsu}},
  \bibinfo {author} {\bibnamefont {B.Zeng}}, \bibinfo {author} {\bibfnamefont
  {X.}~\bibnamefont {Chen}}, \bibinfo {author} {\bibfnamefont {M.~C.}\
  \bibnamefont {Hatnean}}, \bibinfo {author} {\bibfnamefont {Q.~R.}\
  \bibnamefont {Zhang}}, \bibinfo {author} {\bibfnamefont {S.}~\bibnamefont
  {Nakamura}}, \bibinfo {author} {\bibfnamefont {A.~S.}\ \bibnamefont
  {Padgett}}, \bibinfo {author} {\bibfnamefont {G.}~\bibnamefont
  {Rodway-Gant}}, \bibinfo {author} {\bibfnamefont {J.}~\bibnamefont {Berk}},
  \bibinfo {author} {\bibfnamefont {M.~K.}\ \bibnamefont {Kingston}}, \bibinfo
  {author} {\bibfnamefont {G.~H.}\ \bibnamefont {Zhang}}, \bibinfo {author}
  {\bibfnamefont {M.~K.}\ \bibnamefont {Chan}}, \bibinfo {author}
  {\bibfnamefont {S.}~\bibnamefont {Yamashita}}, \bibinfo {author}
  {\bibfnamefont {T.}~\bibnamefont {Sakakibara}}, \bibinfo {author}
  {\bibfnamefont {Y.}~\bibnamefont {Tanako}}, \bibinfo {author} {\bibfnamefont
  {J.-H.}\ \bibnamefont {Park}}, \bibinfo {author} {\bibfnamefont
  {L.}~\bibnamefont {Balicas}}, \bibinfo {author} {\bibfnamefont
  {N.}~\bibnamefont {Harrison}}, \bibinfo {author} {\bibfnamefont
  {N.}~\bibnamefont {Shitsevalova}}, \bibinfo {author} {\bibfnamefont
  {G.}~\bibnamefont {Balakrishnan}}, \bibinfo {author} {\bibfnamefont {G.~G.}\
  \bibnamefont {Lonzarich}}, \bibinfo {author} {\bibfnamefont {R.~W.}\
  \bibnamefont {Hill}}, \bibinfo {author} {\bibfnamefont {M.}~\bibnamefont
  {Sutherland}}, \ and\ \bibinfo {author} {\bibfnamefont {S.~E.}\ \bibnamefont
  {Sebastian}},\ }\href@noop {} {\bibfield  {journal} {\bibinfo  {journal}
  {Nature Physics}\ }\textbf {\bibinfo {volume} {14}},\ \bibinfo {pages} {166}
  (\bibinfo {year} {2017})}\BibitemShut {NoStop}%
\bibitem [{\citenamefont {Thomas}\ \emph {et~al.}(2019)\citenamefont {Thomas},
  \citenamefont {Ding}, \citenamefont {Ronning}, \citenamefont {Zapf},
  \citenamefont {Thompson}, \citenamefont {Fisk}, \citenamefont {Xia},\ and\
  \citenamefont {Rosa}}]{thomas19}%
  \BibitemOpen
  \bibfield  {author} {\bibinfo {author} {\bibfnamefont {S.~M.}\ \bibnamefont
  {Thomas}}, \bibinfo {author} {\bibfnamefont {X.}~\bibnamefont {Ding}},
  \bibinfo {author} {\bibfnamefont {F.}~\bibnamefont {Ronning}}, \bibinfo
  {author} {\bibfnamefont {V.}~\bibnamefont {Zapf}}, \bibinfo {author}
  {\bibfnamefont {J.~D.}\ \bibnamefont {Thompson}}, \bibinfo {author}
  {\bibfnamefont {Z.}~\bibnamefont {Fisk}}, \bibinfo {author} {\bibfnamefont
  {J.}~\bibnamefont {Xia}}, \ and\ \bibinfo {author} {\bibfnamefont {P.~F.~S.}\
  \bibnamefont {Rosa}},\ }\href@noop {} {\bibfield  {journal} {\bibinfo
  {journal} {Physical Review Letters}\ }\textbf {\bibinfo {volume} {122}},\
  \bibinfo {pages} {166401} (\bibinfo {year} {2019})}\BibitemShut {NoStop}%
\bibitem [{\citenamefont {Xu}\ \emph {et~al.}(2016)\citenamefont {Xu},
  \citenamefont {Cui}, \citenamefont {Dong}, \citenamefont {Zhao},
  \citenamefont {Wu}, \citenamefont {Chen}, \citenamefont {Sun}, \citenamefont
  {Yao},\ and\ \citenamefont {Li}}]{xu16}%
  \BibitemOpen
  \bibfield  {author} {\bibinfo {author} {\bibfnamefont {Y.}~\bibnamefont
  {Xu}}, \bibinfo {author} {\bibfnamefont {S.}~\bibnamefont {Cui}}, \bibinfo
  {author} {\bibfnamefont {J.~K.}\ \bibnamefont {Dong}}, \bibinfo {author}
  {\bibfnamefont {D.}~\bibnamefont {Zhao}}, \bibinfo {author} {\bibfnamefont
  {T.}~\bibnamefont {Wu}}, \bibinfo {author} {\bibfnamefont {X.~H.}\
  \bibnamefont {Chen}}, \bibinfo {author} {\bibfnamefont {K.}~\bibnamefont
  {Sun}}, \bibinfo {author} {\bibfnamefont {H.}~\bibnamefont {Yao}}, \ and\
  \bibinfo {author} {\bibfnamefont {S.~Y.}\ \bibnamefont {Li}},\ }\href@noop {}
  {\bibfield  {journal} {\bibinfo  {journal} {Physical Review Letters}\
  }\textbf {\bibinfo {volume} {116}},\ \bibinfo {pages} {246403} (\bibinfo
  {year} {2016})}\BibitemShut {NoStop}%
\bibitem [{\citenamefont {Boulanger}\ \emph {et~al.}(2018)\citenamefont
  {Boulanger}, \citenamefont {Lalibert\'{e}}, \citenamefont {Dion},
  \citenamefont {Badoux}, \citenamefont {Doiron-Leyraud}, \citenamefont
  {Phelan}, \citenamefont {Koohpayeh}, \citenamefont {Fuhrman}, \citenamefont
  {Chamorro}, \citenamefont {McQueen}, \citenamefont {Wang}, \citenamefont
  {Nakajima}, \citenamefont {Metz}, \citenamefont {Paglione},\ and\
  \citenamefont {Taillefer}}]{boulanger}%
  \BibitemOpen
  \bibfield  {author} {\bibinfo {author} {\bibfnamefont {M.-E.}\ \bibnamefont
  {Boulanger}}, \bibinfo {author} {\bibfnamefont {F.}~\bibnamefont
  {Lalibert\'{e}}}, \bibinfo {author} {\bibfnamefont {M.}~\bibnamefont {Dion}},
  \bibinfo {author} {\bibfnamefont {S.}~\bibnamefont {Badoux}}, \bibinfo
  {author} {\bibfnamefont {N.}~\bibnamefont {Doiron-Leyraud}}, \bibinfo
  {author} {\bibfnamefont {W.~A.}\ \bibnamefont {Phelan}}, \bibinfo {author}
  {\bibfnamefont {S.~M.}\ \bibnamefont {Koohpayeh}}, \bibinfo {author}
  {\bibfnamefont {W.~T.}\ \bibnamefont {Fuhrman}}, \bibinfo {author}
  {\bibfnamefont {J.~R.}\ \bibnamefont {Chamorro}}, \bibinfo {author}
  {\bibfnamefont {T.~M.}\ \bibnamefont {McQueen}}, \bibinfo {author}
  {\bibfnamefont {X.~F.}\ \bibnamefont {Wang}}, \bibinfo {author}
  {\bibfnamefont {Y.}~\bibnamefont {Nakajima}}, \bibinfo {author}
  {\bibfnamefont {T.}~\bibnamefont {Metz}}, \bibinfo {author} {\bibfnamefont
  {J.}~\bibnamefont {Paglione}}, \ and\ \bibinfo {author} {\bibfnamefont
  {L.}~\bibnamefont {Taillefer}},\ }\href@noop {} {\bibfield  {journal}
  {\bibinfo  {journal} {Physical Review B}\ }\textbf {\bibinfo {volume} {97}},\
  \bibinfo {pages} {245141} (\bibinfo {year} {2018})}\BibitemShut {NoStop}%
\bibitem [{\citenamefont {{J. Knolle and N. R. Cooper}}(2017)}]{knolle17}%
  \BibitemOpen
  \bibfield  {author} {\bibinfo {author} {\bibnamefont {{J. Knolle and N. R.
  Cooper}}},\ }\href@noop {} {\bibfield  {journal} {\bibinfo  {journal}
  {Physical Review Letters}\ }\textbf {\bibinfo {volume} {118}},\ \bibinfo
  {pages} {096604} (\bibinfo {year} {2017})}\BibitemShut {NoStop}%
\bibitem [{\citenamefont {Chowdhury}\ \emph {et~al.}(2018)\citenamefont
  {Chowdhury}, \citenamefont {Sodemann},\ and\ \citenamefont
  {Senthil}}]{chowdhury}%
  \BibitemOpen
  \bibfield  {author} {\bibinfo {author} {\bibfnamefont {D.}~\bibnamefont
  {Chowdhury}}, \bibinfo {author} {\bibfnamefont {I.}~\bibnamefont {Sodemann}},
  \ and\ \bibinfo {author} {\bibfnamefont {T.}~\bibnamefont {Senthil}},\
  }\href@noop {} {\bibfield  {journal} {\bibinfo  {journal} {Nature
  Communications}\ }\textbf {\bibinfo {volume} {9}},\ \bibinfo {pages} {1766}
  (\bibinfo {year} {2018})}\BibitemShut {NoStop}%
\bibitem [{\citenamefont {Baskaran}(2015)}]{baskaran}%
  \BibitemOpen
  \bibfield  {author} {\bibinfo {author} {\bibfnamefont {G.}~\bibnamefont
  {Baskaran}},\ }\href@noop {} {\enquote {\bibinfo {title} {{Majorana Fermi sea
  in insulating SmB$_6$: A proposal and a theory of quantum oscillations in
  Kondo insulators}},}\ } (\bibinfo {year} {2015}),\ \bibinfo {note}
  {{arXiv:1507.03477v1 [cond-mat.str-el]}}\BibitemShut {NoStop}%
\bibitem [{\citenamefont {Erten}\ \emph
  {et~al.}(2016{\natexlab{a}})\citenamefont {Erten}, \citenamefont {Ghaemi},\
  and\ \citenamefont {Coleman}}]{erten17}%
  \BibitemOpen
  \bibfield  {author} {\bibinfo {author} {\bibfnamefont {O.}~\bibnamefont
  {Erten}}, \bibinfo {author} {\bibfnamefont {P.}~\bibnamefont {Ghaemi}}, \
  and\ \bibinfo {author} {\bibfnamefont {P.}~\bibnamefont {Coleman}},\
  }\href@noop {} {\bibfield  {journal} {\bibinfo  {journal} {Physical Review
  Letters}\ }\textbf {\bibinfo {volume} {116}},\ \bibinfo {pages} {046403}
  (\bibinfo {year} {2016}{\natexlab{a}})}\BibitemShut {NoStop}%
\bibitem [{\citenamefont {Erten}\ \emph
  {et~al.}(2016{\natexlab{b}})\citenamefont {Erten}, \citenamefont {Ghaemi},\
  and\ \citenamefont {Coleman}}]{erten16}%
  \BibitemOpen
  \bibfield  {author} {\bibinfo {author} {\bibfnamefont {O.}~\bibnamefont
  {Erten}}, \bibinfo {author} {\bibfnamefont {P.}~\bibnamefont {Ghaemi}}, \
  and\ \bibinfo {author} {\bibfnamefont {P.}~\bibnamefont {Coleman}},\
  }\href@noop {} {\bibfield  {journal} {\bibinfo  {journal} {Physical Review
  Letters}\ }\textbf {\bibinfo {volume} {116}},\ \bibinfo {pages} {046403}
  (\bibinfo {year} {2016}{\natexlab{b}})}\BibitemShut {NoStop}%
\bibitem [{\citenamefont {{P. S. Riseborough and Z.
  Fisk}}(2017)}]{riseborough17}%
  \BibitemOpen
  \bibfield  {author} {\bibinfo {author} {\bibnamefont {{P. S. Riseborough and
  Z. Fisk}}},\ }\href@noop {} {\bibfield  {journal} {\bibinfo  {journal}
  {Physical Review B}\ }\textbf {\bibinfo {volume} {96}},\ \bibinfo {pages}
  {195122} (\bibinfo {year} {2017})}\BibitemShut {NoStop}%
\bibitem [{\citenamefont {{J. Knolle and N. R. Cooper}}(2015)}]{knolle15}%
  \BibitemOpen
  \bibfield  {author} {\bibinfo {author} {\bibnamefont {{J. Knolle and N. R.
  Cooper}}},\ }\href@noop {} {\bibfield  {journal} {\bibinfo  {journal}
  {Physical Review Letters}\ }\textbf {\bibinfo {volume} {115}},\ \bibinfo
  {pages} {146401} (\bibinfo {year} {2015})}\BibitemShut {NoStop}%
\bibitem [{\citenamefont {Zhang}\ \emph {et~al.}(2016)\citenamefont {Zhang},
  \citenamefont {Song},\ and\ \citenamefont {Wang}}]{zhang16}%
  \BibitemOpen
  \bibfield  {author} {\bibinfo {author} {\bibfnamefont {L.}~\bibnamefont
  {Zhang}}, \bibinfo {author} {\bibfnamefont {X.~Y.}\ \bibnamefont {Song}}, \
  and\ \bibinfo {author} {\bibfnamefont {F.}~\bibnamefont {Wang}},\ }\href@noop
  {} {\bibfield  {journal} {\bibinfo  {journal} {Physical Review Letters}\
  }\textbf {\bibinfo {volume} {116}},\ \bibinfo {pages} {046404} (\bibinfo
  {year} {2016})}\BibitemShut {NoStop}%
\bibitem [{\citenamefont {{H. Shen and L. Fu}}(2018)}]{shen}%
  \BibitemOpen
  \bibfield  {author} {\bibinfo {author} {\bibnamefont {{H. Shen and L. Fu}}},\
  }\href@noop {} {\bibfield  {journal} {\bibinfo  {journal} {Physical Review
  Letters}\ }\textbf {\bibinfo {volume} {121}},\ \bibinfo {pages} {026403}
  (\bibinfo {year} {2018})}\BibitemShut {NoStop}%
\bibitem [{\citenamefont {Harrison}(2018)}]{harrison}%
  \BibitemOpen
  \bibfield  {author} {\bibinfo {author} {\bibfnamefont {N.}~\bibnamefont
  {Harrison}},\ }\href@noop {} {\bibfield  {journal} {\bibinfo  {journal}
  {Physical Review Letters}\ }\textbf {\bibinfo {volume} {121}},\ \bibinfo
  {pages} {026602} (\bibinfo {year} {2018})}\BibitemShut {NoStop}%
\bibitem [{\citenamefont {Fuhrman}\ \emph {et~al.}(2018)\citenamefont
  {Fuhrman}, \citenamefont {Chamorro}, \citenamefont {Alekseev}, \citenamefont
  {Mignot}, \citenamefont {Keller}, \citenamefont {{Rodriguez-Rivera}},
  \citenamefont {Qiu}, \citenamefont {McQueen},\ and\ \citenamefont
  {Broholm}}]{fuhrman18}%
  \BibitemOpen
  \bibfield  {author} {\bibinfo {author} {\bibfnamefont {W.~T.}\ \bibnamefont
  {Fuhrman}}, \bibinfo {author} {\bibfnamefont {J.~R.}\ \bibnamefont
  {Chamorro}}, \bibinfo {author} {\bibfnamefont {P.~A.}\ \bibnamefont
  {Alekseev}}, \bibinfo {author} {\bibfnamefont {J.-M.}\ \bibnamefont
  {Mignot}}, \bibinfo {author} {\bibfnamefont {T.}~\bibnamefont {Keller}},
  \bibinfo {author} {\bibfnamefont {J.~A.}\ \bibnamefont {{Rodriguez-Rivera}}},
  \bibinfo {author} {\bibfnamefont {Y.}~\bibnamefont {Qiu}}, \bibinfo {author}
  {\bibfnamefont {P.~N. T.~M.}\ \bibnamefont {McQueen}}, \ and\ \bibinfo
  {author} {\bibfnamefont {C.}~\bibnamefont {Broholm}},\ }\href@noop {}
  {\bibfield  {journal} {\bibinfo  {journal} {Nature Communications}\ }\textbf
  {\bibinfo {volume} {9}},\ \bibinfo {pages} {1539} (\bibinfo {year}
  {2018})}\BibitemShut {NoStop}%
\bibitem [{\citenamefont {{W. T. Fuhrman and P.
  Nikoli\'{c}}}(2018)}]{fuhrman182}%
  \BibitemOpen
  \bibfield  {author} {\bibinfo {author} {\bibnamefont {{W. T. Fuhrman and P.
  Nikoli\'{c}}}},\ }\href@noop {} {\enquote {\bibinfo {title} {{Magnetic
  impurities in Kondo insulators and the puzzle of samarium hexaboride}},}\ }
  (\bibinfo {year} {2018}),\ \bibinfo {note} {{arXiv:1807.00005v1
  [cond-mat.str-el]}}\BibitemShut {NoStop}%
\bibitem [{\citenamefont {Abele}\ \emph {et~al.}(2020)\citenamefont {Abele},
  \citenamefont {Yuan},\ and\ \citenamefont {Riseborough}}]{abele}%
  \BibitemOpen
  \bibfield  {author} {\bibinfo {author} {\bibfnamefont {M.}~\bibnamefont
  {Abele}}, \bibinfo {author} {\bibfnamefont {X.}~\bibnamefont {Yuan}}, \ and\
  \bibinfo {author} {\bibfnamefont {P.~S.}\ \bibnamefont {Riseborough}},\
  }\href@noop {} {\bibfield  {journal} {\bibinfo  {journal} {Physical Review
  B}\ }\textbf {\bibinfo {volume} {101}},\ \bibinfo {pages} {094101} (\bibinfo
  {year} {2020})}\BibitemShut {NoStop}%
\bibitem [{\citenamefont {Skinner}(2019)}]{skinner}%
  \BibitemOpen
  \bibfield  {author} {\bibinfo {author} {\bibfnamefont {B.}~\bibnamefont
  {Skinner}},\ }\href@noop {} {\bibfield  {journal} {\bibinfo  {journal}
  {Physical Review Materials}\ }\textbf {\bibinfo {volume} {3}},\ \bibinfo
  {pages} {104601} (\bibinfo {year} {2019})}\BibitemShut {NoStop}%
\bibitem [{\citenamefont {Rakoski}\ \emph {et~al.}(2017)\citenamefont
  {Rakoski}, \citenamefont {Eo}, \citenamefont {Sun},\ and\ \citenamefont
  {\c{C}. Kurdak}}]{rakoski17}%
  \BibitemOpen
  \bibfield  {author} {\bibinfo {author} {\bibfnamefont {A.}~\bibnamefont
  {Rakoski}}, \bibinfo {author} {\bibfnamefont {Y.~S.}\ \bibnamefont {Eo}},
  \bibinfo {author} {\bibfnamefont {K.}~\bibnamefont {Sun}}, \ and\ \bibinfo
  {author} {\bibnamefont {\c{C}. Kurdak}},\ }\href@noop {} {\bibfield
  {journal} {\bibinfo  {journal} {Physical Review B}\ }\textbf {\bibinfo
  {volume} {95}},\ \bibinfo {pages} {195133} (\bibinfo {year}
  {2017})}\BibitemShut {NoStop}%
\bibitem [{\citenamefont {Eo}\ \emph {et~al.}(2018)\citenamefont {Eo},
  \citenamefont {Sun}, \citenamefont {\c{C}. Kurdak}, \citenamefont {Kim},\
  and\ \citenamefont {Fisk}}]{eo18}%
  \BibitemOpen
  \bibfield  {author} {\bibinfo {author} {\bibfnamefont {Y.~S.}\ \bibnamefont
  {Eo}}, \bibinfo {author} {\bibfnamefont {K.}~\bibnamefont {Sun}}, \bibinfo
  {author} {\bibnamefont {\c{C}. Kurdak}}, \bibinfo {author} {\bibfnamefont
  {D.-J.}\ \bibnamefont {Kim}}, \ and\ \bibinfo {author} {\bibfnamefont
  {Z.}~\bibnamefont {Fisk}},\ }\href@noop {} {\bibfield  {journal} {\bibinfo
  {journal} {Physical Review Applied}\ }\textbf {\bibinfo {volume} {9}},\
  \bibinfo {pages} {044006} (\bibinfo {year} {2018})}\BibitemShut {NoStop}%
\bibitem [{\citenamefont {Eo}\ \emph {et~al.}(2019)\citenamefont {Eo},
  \citenamefont {Rakoski}, \citenamefont {Lucien}, \citenamefont {Mihaliov},
  \citenamefont {\c{C}. Kurdak}, \citenamefont {Rosa},\ and\ \citenamefont
  {Fisk}}]{eo19}%
  \BibitemOpen
  \bibfield  {author} {\bibinfo {author} {\bibfnamefont {Y.~S.}\ \bibnamefont
  {Eo}}, \bibinfo {author} {\bibfnamefont {A.}~\bibnamefont {Rakoski}},
  \bibinfo {author} {\bibfnamefont {J.}~\bibnamefont {Lucien}}, \bibinfo
  {author} {\bibfnamefont {D.}~\bibnamefont {Mihaliov}}, \bibinfo {author}
  {\bibnamefont {\c{C}. Kurdak}}, \bibinfo {author} {\bibfnamefont {P.~F.~S.}\
  \bibnamefont {Rosa}}, \ and\ \bibinfo {author} {\bibfnamefont
  {Z.}~\bibnamefont {Fisk}},\ }\href@noop {} {\bibfield  {journal} {\bibinfo
  {journal} {Proceedings of the National Academy of Sciences}\ }\textbf
  {\bibinfo {volume} {116}},\ \bibinfo {pages} {12638} (\bibinfo {year}
  {2019})}\BibitemShut {NoStop}%
\bibitem [{\citenamefont {Debye}\ and\ \citenamefont {Conwell}(1954)}]{debye}%
  \BibitemOpen
  \bibfield  {author} {\bibinfo {author} {\bibfnamefont {P.~P.}\ \bibnamefont
  {Debye}}\ and\ \bibinfo {author} {\bibfnamefont {E.~M.}\ \bibnamefont
  {Conwell}},\ }\href@noop {} {\bibfield  {journal} {\bibinfo  {journal}
  {Physical Review}\ }\textbf {\bibinfo {volume} {93}},\ \bibinfo {pages} {693}
  (\bibinfo {year} {1954})}\BibitemShut {NoStop}%
\bibitem [{\citenamefont {Chapman}\ \emph {et~al.}(1963)\citenamefont
  {Chapman}, \citenamefont {Tufte}, \citenamefont {Zook},\ and\ \citenamefont
  {Long}}]{chapman}%
  \BibitemOpen
  \bibfield  {author} {\bibinfo {author} {\bibfnamefont {P.~W.}\ \bibnamefont
  {Chapman}}, \bibinfo {author} {\bibfnamefont {O.~N.}\ \bibnamefont {Tufte}},
  \bibinfo {author} {\bibfnamefont {J.~D.}\ \bibnamefont {Zook}}, \ and\
  \bibinfo {author} {\bibfnamefont {D.}~\bibnamefont {Long}},\ }\href@noop {}
  {\bibfield  {journal} {\bibinfo  {journal} {Journal of Applied Physics}\
  }\textbf {\bibinfo {volume} {34}},\ \bibinfo {pages} {3291} (\bibinfo {year}
  {1963})}\BibitemShut {NoStop}%
\bibitem [{\citenamefont {Ourmazd}(1984)}]{ourmazd}%
  \BibitemOpen
  \bibfield  {author} {\bibinfo {author} {\bibfnamefont {A.}~\bibnamefont
  {Ourmazd}},\ }\href@noop {} {\bibfield  {journal} {\bibinfo  {journal}
  {Contemporary Physics}\ }\textbf {\bibinfo {volume} {25}},\ \bibinfo {pages}
  {251} (\bibinfo {year} {1984})}\BibitemShut {NoStop}%
\bibitem [{\citenamefont {Hall}\ and\ \citenamefont {Bean}(1992)}]{hullbean}%
  \BibitemOpen
  \bibfield  {author} {\bibinfo {author} {\bibfnamefont {R.}~\bibnamefont
  {Hall}}\ and\ \bibinfo {author} {\bibfnamefont {J.~C.}\ \bibnamefont
  {Bean}},\ }\href@noop {} {\bibfield  {journal} {\bibinfo  {journal} {Critical
  Reviews in Solid State and Material Sciences}\ }\textbf {\bibinfo {volume}
  {17}},\ \bibinfo {pages} {507} (\bibinfo {year} {1992})}\BibitemShut
  {NoStop}%
\bibitem [{\citenamefont {{D. Hull and D. J. Bacon}}(2011)}]{hull}%
  \BibitemOpen
  \bibfield  {author} {\bibinfo {author} {\bibnamefont {{D. Hull and D. J.
  Bacon}}},\ }\href@noop {} {\emph {\bibinfo {title} {Introduction to
  Dislocations}}},\ \bibinfo {edition} {5th}\ ed.\ (\bibinfo  {publisher}
  {Elsevier, Ltd.},\ \bibinfo {year} {2011})\BibitemShut {NoStop}%
\bibitem [{\citenamefont {Hamasaki}\ \emph {et~al.}(2017)\citenamefont
  {Hamasaki}, \citenamefont {Tokumoto},\ and\ \citenamefont
  {Edagawa}}]{hamasaki}%
  \BibitemOpen
  \bibfield  {author} {\bibinfo {author} {\bibfnamefont {H.}~\bibnamefont
  {Hamasaki}}, \bibinfo {author} {\bibfnamefont {Y.}~\bibnamefont {Tokumoto}},
  \ and\ \bibinfo {author} {\bibfnamefont {K.}~\bibnamefont {Edagawa}},\
  }\href@noop {} {\bibfield  {journal} {\bibinfo  {journal} {Applied Physics
  Letters}\ }\textbf {\bibinfo {volume} {110}},\ \bibinfo {pages} {092105}
  (\bibinfo {year} {2017})}\BibitemShut {NoStop}%
\bibitem [{\citenamefont {Ran}\ \emph {et~al.}(2009)\citenamefont {Ran},
  \citenamefont {Zhang},\ and\ \citenamefont {Vishwanath}}]{ran}%
  \BibitemOpen
  \bibfield  {author} {\bibinfo {author} {\bibfnamefont {Y.}~\bibnamefont
  {Ran}}, \bibinfo {author} {\bibfnamefont {Y.}~\bibnamefont {Zhang}}, \ and\
  \bibinfo {author} {\bibfnamefont {A.}~\bibnamefont {Vishwanath}},\
  }\href@noop {} {\bibfield  {journal} {\bibinfo  {journal} {Nature Physics}\
  }\textbf {\bibinfo {volume} {5}},\ \bibinfo {pages} {298} (\bibinfo {year}
  {2009})}\BibitemShut {NoStop}%
\bibitem [{\citenamefont {Luo}\ \emph {et~al.}(2015)\citenamefont {Luo},
  \citenamefont {Chen}, \citenamefont {Dai}, \citenamefont {Xu},\ and\
  \citenamefont {Thompson}}]{luo15}%
  \BibitemOpen
  \bibfield  {author} {\bibinfo {author} {\bibfnamefont {Y.}~\bibnamefont
  {Luo}}, \bibinfo {author} {\bibfnamefont {H.}~\bibnamefont {Chen}}, \bibinfo
  {author} {\bibfnamefont {J.}~\bibnamefont {Dai}}, \bibinfo {author}
  {\bibfnamefont {Z.}~\bibnamefont {Xu}}, \ and\ \bibinfo {author}
  {\bibfnamefont {J.~D.}\ \bibnamefont {Thompson}},\ }\href@noop {} {\bibfield
  {journal} {\bibinfo  {journal} {Physical Review B}\ }\textbf {\bibinfo
  {volume} {91}},\ \bibinfo {pages} {075130} (\bibinfo {year}
  {2015})}\BibitemShut {NoStop}%
\bibitem [{\citenamefont {Eo}\ \emph {et~al.}(2020)\citenamefont {Eo},
  \citenamefont {Wolgast}, \citenamefont {Rakoski}, \citenamefont {Mihaliov},
  \citenamefont {Kang}, \citenamefont {Song}, \citenamefont {Cho},
  \citenamefont {Hatnean}, \citenamefont {Balakrishnan}, \citenamefont {Fisk},
  \citenamefont {Saha}, \citenamefont {Wang}, \citenamefont {Paglione},\ and\
  \citenamefont {\c{C}. Kurdak}}]{eo20}%
  \BibitemOpen
  \bibfield  {author} {\bibinfo {author} {\bibfnamefont {Y.~S.}\ \bibnamefont
  {Eo}}, \bibinfo {author} {\bibfnamefont {S.}~\bibnamefont {Wolgast}},
  \bibinfo {author} {\bibfnamefont {A.}~\bibnamefont {Rakoski}}, \bibinfo
  {author} {\bibfnamefont {D.}~\bibnamefont {Mihaliov}}, \bibinfo {author}
  {\bibfnamefont {B.~Y.}\ \bibnamefont {Kang}}, \bibinfo {author}
  {\bibfnamefont {M.~S.}\ \bibnamefont {Song}}, \bibinfo {author}
  {\bibfnamefont {B.~K.}\ \bibnamefont {Cho}}, \bibinfo {author} {\bibfnamefont
  {M.~C.}\ \bibnamefont {Hatnean}}, \bibinfo {author} {\bibfnamefont
  {G.}~\bibnamefont {Balakrishnan}}, \bibinfo {author} {\bibfnamefont
  {Z.}~\bibnamefont {Fisk}}, \bibinfo {author} {\bibfnamefont {S.~R.}\
  \bibnamefont {Saha}}, \bibinfo {author} {\bibfnamefont {X.}~\bibnamefont
  {Wang}}, \bibinfo {author} {\bibfnamefont {J.}~\bibnamefont {Paglione}}, \
  and\ \bibinfo {author} {\bibnamefont {\c{C}. Kurdak}},\ }\href@noop {}
  {\bibfield  {journal} {\bibinfo  {journal} {Physical Review B}\ }\textbf
  {\bibinfo {volume} {101}},\ \bibinfo {pages} {155109} (\bibinfo {year}
  {2020})}\BibitemShut {NoStop}%
\bibitem [{\citenamefont {Cooley}\ \emph {et~al.}(1995)\citenamefont {Cooley},
  \citenamefont {Aronson}, \citenamefont {Fisk},\ and\ \citenamefont
  {Canfield}}]{cooleyprl}%
  \BibitemOpen
  \bibfield  {author} {\bibinfo {author} {\bibfnamefont {J.~C.}\ \bibnamefont
  {Cooley}}, \bibinfo {author} {\bibfnamefont {M.~C.}\ \bibnamefont {Aronson}},
  \bibinfo {author} {\bibfnamefont {Z.}~\bibnamefont {Fisk}}, \ and\ \bibinfo
  {author} {\bibfnamefont {P.~C.}\ \bibnamefont {Canfield}},\ }\href@noop {}
  {\bibfield  {journal} {\bibinfo  {journal} {Physical Review Letters}\
  }\textbf {\bibinfo {volume} {74}},\ \bibinfo {pages} {1629} (\bibinfo {year}
  {1995})}\BibitemShut {NoStop}%
\bibitem [{\citenamefont {Sluchanko}\ \emph {et~al.}(1999)\citenamefont
  {Sluchanko}, \citenamefont {Volkov}, \citenamefont {Glushkov}, \citenamefont
  {Gorshunov}, \citenamefont {Demishev}, \citenamefont {Kondrin}, \citenamefont
  {Pronin}, \citenamefont {Samarin}, \citenamefont {Bruynseraede},
  \citenamefont {Moshchalkov},\ and\ \citenamefont {Kunii}}]{sluchanko99}%
  \BibitemOpen
  \bibfield  {author} {\bibinfo {author} {\bibfnamefont {N.~E.}\ \bibnamefont
  {Sluchanko}}, \bibinfo {author} {\bibfnamefont {A.~A.}\ \bibnamefont
  {Volkov}}, \bibinfo {author} {\bibfnamefont {V.~V.}\ \bibnamefont
  {Glushkov}}, \bibinfo {author} {\bibfnamefont {B.~P.}\ \bibnamefont
  {Gorshunov}}, \bibinfo {author} {\bibfnamefont {S.~V.}\ \bibnamefont
  {Demishev}}, \bibinfo {author} {\bibfnamefont {M.~V.}\ \bibnamefont
  {Kondrin}}, \bibinfo {author} {\bibfnamefont {A.~A.}\ \bibnamefont {Pronin}},
  \bibinfo {author} {\bibfnamefont {N.~A.}\ \bibnamefont {Samarin}}, \bibinfo
  {author} {\bibfnamefont {Y.}~\bibnamefont {Bruynseraede}}, \bibinfo {author}
  {\bibfnamefont {V.~V.}\ \bibnamefont {Moshchalkov}}, \ and\ \bibinfo {author}
  {\bibfnamefont {S.}~\bibnamefont {Kunii}},\ }\href@noop {} {\bibfield
  {journal} {\bibinfo  {journal} {Journal of Experimental and Theoretical
  Physics}\ }\textbf {\bibinfo {volume} {88}},\ \bibinfo {pages} {533}
  (\bibinfo {year} {1999})}\BibitemShut {NoStop}%
\bibitem [{\citenamefont {Flachbart}\ \emph {et~al.}(2001)\citenamefont
  {Flachbart}, \citenamefont {Gab\'{a}ni}, \citenamefont {Konovalova},
  \citenamefont {Paderno},\ and\ \citenamefont {Pavlik}}]{flachbart012}%
  \BibitemOpen
  \bibfield  {author} {\bibinfo {author} {\bibfnamefont {K.}~\bibnamefont
  {Flachbart}}, \bibinfo {author} {\bibfnamefont {S.}~\bibnamefont
  {Gab\'{a}ni}}, \bibinfo {author} {\bibfnamefont {E.}~\bibnamefont
  {Konovalova}}, \bibinfo {author} {\bibfnamefont {Y.}~\bibnamefont {Paderno}},
  \ and\ \bibinfo {author} {\bibfnamefont {V.}~\bibnamefont {Pavlik}},\
  }\href@noop {} {\bibfield  {journal} {\bibinfo  {journal} {Physica B}\
  }\textbf {\bibinfo {volume} {293}},\ \bibinfo {pages} {417} (\bibinfo {year}
  {2001})}\BibitemShut {NoStop}%
\bibitem [{\citenamefont {Gab\'{a}ni}\ \emph {et~al.}(2015)\citenamefont
  {Gab\'{a}ni}, \citenamefont {Prist\'{a}s}, \citenamefont
  {Taka\'{a}\v{c}ov\'{a}}, \citenamefont {Sluchanko}, \citenamefont
  {Siemensmeyer}, \citenamefont {Shitsevalova}, \citenamefont {Filipov},\ and\
  \citenamefont {Flachbart}}]{gabani15}%
  \BibitemOpen
  \bibfield  {author} {\bibinfo {author} {\bibfnamefont {S.}~\bibnamefont
  {Gab\'{a}ni}}, \bibinfo {author} {\bibfnamefont {G.}~\bibnamefont
  {Prist\'{a}s}}, \bibinfo {author} {\bibfnamefont {I.}~\bibnamefont
  {Taka\'{a}\v{c}ov\'{a}}}, \bibinfo {author} {\bibfnamefont {N.}~\bibnamefont
  {Sluchanko}}, \bibinfo {author} {\bibfnamefont {K.}~\bibnamefont
  {Siemensmeyer}}, \bibinfo {author} {\bibfnamefont {N.}~\bibnamefont
  {Shitsevalova}}, \bibinfo {author} {\bibfnamefont {V.}~\bibnamefont
  {Filipov}}, \ and\ \bibinfo {author} {\bibfnamefont {K.}~\bibnamefont
  {Flachbart}},\ }\href@noop {} {\bibfield  {journal} {\bibinfo  {journal}
  {Solid State Sciences}\ }\textbf {\bibinfo {volume} {47}},\ \bibinfo {pages}
  {17} (\bibinfo {year} {2015})}\BibitemShut {NoStop}%
\bibitem [{\citenamefont {Geballe}\ \emph {et~al.}(1970)\citenamefont
  {Geballe}, \citenamefont {Menth}, \citenamefont {Buehler},\ and\
  \citenamefont {Hull}}]{geballe}%
  \BibitemOpen
  \bibfield  {author} {\bibinfo {author} {\bibfnamefont {T.~H.}\ \bibnamefont
  {Geballe}}, \bibinfo {author} {\bibfnamefont {A.}~\bibnamefont {Menth}},
  \bibinfo {author} {\bibfnamefont {E.}~\bibnamefont {Buehler}}, \ and\
  \bibinfo {author} {\bibfnamefont {G.~W.}\ \bibnamefont {Hull}},\ }\href@noop
  {} {\bibfield  {journal} {\bibinfo  {journal} {Journal of Applied Physics}\
  }\textbf {\bibinfo {volume} {41}},\ \bibinfo {pages} {904} (\bibinfo {year}
  {1970})}\BibitemShut {NoStop}%
\bibitem [{\citenamefont {Fuhrman}\ \emph {et~al.}(2019)\citenamefont
  {Fuhrman}, \citenamefont {Leiner}, \citenamefont {Freeland}, \citenamefont
  {{van Veenendaal}}, \citenamefont {Koohpayeh}, \citenamefont {Phelan},
  \citenamefont {McQueen},\ and\ \citenamefont {Broholm}}]{fuhrman19}%
  \BibitemOpen
  \bibfield  {author} {\bibinfo {author} {\bibfnamefont {W.~T.}\ \bibnamefont
  {Fuhrman}}, \bibinfo {author} {\bibfnamefont {J.~C.}\ \bibnamefont {Leiner}},
  \bibinfo {author} {\bibfnamefont {J.~W.}\ \bibnamefont {Freeland}}, \bibinfo
  {author} {\bibfnamefont {M.}~\bibnamefont {{van Veenendaal}}}, \bibinfo
  {author} {\bibfnamefont {S.~M.}\ \bibnamefont {Koohpayeh}}, \bibinfo {author}
  {\bibfnamefont {W.~A.}\ \bibnamefont {Phelan}}, \bibinfo {author}
  {\bibfnamefont {T.~M.}\ \bibnamefont {McQueen}}, \ and\ \bibinfo {author}
  {\bibfnamefont {C.}~\bibnamefont {Broholm}},\ }\href@noop {} {\bibfield
  {journal} {\bibinfo  {journal} {Physical Review B}\ }\textbf {\bibinfo
  {volume} {99}},\ \bibinfo {pages} {020401(R)} (\bibinfo {year}
  {2019})}\BibitemShut {NoStop}%
\bibitem [{\citenamefont {Hartstein}\ \emph {et~al.}(2020)\citenamefont
  {Hartstein}, \citenamefont {Liu}, \citenamefont {Hsu}, \citenamefont {Tan},
  \citenamefont {Hatnean}, \citenamefont {Balakrishnan},\ and\ \citenamefont
  {Sebastian}}]{hartstein20}%
  \BibitemOpen
  \bibfield  {author} {\bibinfo {author} {\bibfnamefont {M.}~\bibnamefont
  {Hartstein}}, \bibinfo {author} {\bibfnamefont {H.}~\bibnamefont {Liu}},
  \bibinfo {author} {\bibfnamefont {Y.-T.}\ \bibnamefont {Hsu}}, \bibinfo
  {author} {\bibfnamefont {B.~S.}\ \bibnamefont {Tan}}, \bibinfo {author}
  {\bibfnamefont {M.~C.}\ \bibnamefont {Hatnean}}, \bibinfo {author}
  {\bibfnamefont {G.}~\bibnamefont {Balakrishnan}}, \ and\ \bibinfo {author}
  {\bibfnamefont {S.~E.}\ \bibnamefont {Sebastian}},\ }\href@noop {} {\bibfield
   {journal} {\bibinfo  {journal} {iScience}\ }\textbf {\bibinfo {volume}
  {23}},\ \bibinfo {pages} {101632,} (\bibinfo {year} {2020})}\BibitemShut
  {NoStop}%
\bibitem [{\citenamefont {Souza}\ \emph {et~al.}(2020)\citenamefont {Souza},
  \citenamefont {Rosa}, \citenamefont {Sichelschmidt}, \citenamefont {Carlone},
  \citenamefont {Venegas}, \citenamefont {Malcolms}, \citenamefont {Menegasso},
  \citenamefont {Urbano}, \citenamefont {Fisk},\ and\ \citenamefont
  {Pagliuso}}]{souza20}%
  \BibitemOpen
  \bibfield  {author} {\bibinfo {author} {\bibfnamefont {J.~C.}\ \bibnamefont
  {Souza}}, \bibinfo {author} {\bibfnamefont {P.~F.~S.}\ \bibnamefont {Rosa}},
  \bibinfo {author} {\bibfnamefont {J.}~\bibnamefont {Sichelschmidt}}, \bibinfo
  {author} {\bibfnamefont {M.}~\bibnamefont {Carlone}}, \bibinfo {author}
  {\bibfnamefont {P.~A.}\ \bibnamefont {Venegas}}, \bibinfo {author}
  {\bibfnamefont {M.~O.}\ \bibnamefont {Malcolms}}, \bibinfo {author}
  {\bibfnamefont {P.~M.}\ \bibnamefont {Menegasso}}, \bibinfo {author}
  {\bibfnamefont {R.~R.}\ \bibnamefont {Urbano}}, \bibinfo {author}
  {\bibfnamefont {Z.}~\bibnamefont {Fisk}}, \ and\ \bibinfo {author}
  {\bibfnamefont {P.~G.}\ \bibnamefont {Pagliuso}},\ }\href@noop {} {\enquote
  {\bibinfo {title} {{Metallic islands in the Kondo insulator SmB$_6$}},}\ }
  (\bibinfo {year} {2020}),\ \bibinfo {note} {{arXiv:2010.03719
  [cond-mat.str-el]}}\BibitemShut {NoStop}%
\bibitem [{\citenamefont {Kohn}(1957)}]{kohn}%
  \BibitemOpen
  \bibfield  {author} {\bibinfo {author} {\bibfnamefont {W.}~\bibnamefont
  {Kohn}},\ }\href@noop {} {\bibfield  {journal} {\bibinfo  {journal} {Physical
  Review}\ }\textbf {\bibinfo {volume} {105}},\ \bibinfo {pages} {509}
  (\bibinfo {year} {1957})}\BibitemShut {NoStop}%
\end{thebibliography}%

\end{document}